\pdfoutput=1
\documentclass[11pt,a4paper]{article}
\usepackage{jheppub}

\RequirePackage[utf8]{inputenc}

\usepackage{listings}
\usepackage{amsfonts}
\usepackage{amssymb}
\usepackage{comment}
\usepackage{graphicx}
\usepackage{amsmath}
\usepackage{tabu}
\usepackage{dsfont}
\usepackage{float}
\usepackage{amsthm}
\usepackage{slashed}
\usepackage{setspace}
\usepackage[labelformat=simple]{subcaption}

\usepackage[boxsize=0.5em,aligntableaux=center]{ytableau}
\usepackage[utf8]{inputenc}
\usepackage{amsmath, amsthm, amssymb, amsfonts}
\usepackage[font=small, labelfont=bf]{caption}
\usepackage{amsmath, amsthm, amssymb, amsfonts}
\usepackage{multirow}
\usepackage{graphicx}
\usepackage{float}
\usepackage[numbers,sort&compress]{natbib}
\usepackage{jheppub}
\usepackage{enumerate}
\usepackage{amsmath}
\usepackage{amsfonts}
\usepackage{amssymb}
\usepackage{graphicx}
\usepackage{cancel} 
\usepackage{algorithm2e}
\usepackage{empheq}
\usepackage{color}
\usepackage{hyperref}
\usepackage{float}
\restylefloat{table}
\restylefloat{figure}

\newcommand{\nn}{\nonumber}

\newcommand{\nc}{\newcommand}
\nc{\beq}{\begin{equation}}
\nc{\eeq}{\end{equation}}
\nc{\be}{\begin{equation}}
\nc{\ee}{\end{equation}}
\nc{\bea}{\begin{eqnarray}}
\nc{\eea}{\end{eqnarray}}
\nc{\bi}{\begin{itemize}}
\nc{\ei}{\end{itemize}}
\nc{\ben}{\begin{enumerate}}
\nc{\een}{\end{enumerate}}

\def\ov{\overline}

\newcommand{\la}{\langle}
\newcommand{\ra}{\rangle}

\allowdisplaybreaks[2]
\numberwithin{equation}{section}

\def\vo{\mathcal{V}}

\allowdisplaybreaks[2]
\numberwithin{equation}{section}

\lstdefinestyle{customc}{
  belowcaptionskip=1\baselineskip,
  breaklines=true,
  frame=L,
  xleftmargin=\parindent,
  language=C++,
  showstringspaces=false,
  basicstyle=\tiny\ttfamily,
  keywordstyle=\bfseries\color{green!40!black},
  commentstyle=\itshape\color{purple!40!black},
  identifierstyle=\color{blue},
  stringstyle=\color{orange},
}

\title{A Systematic Approach to K\"ahler Moduli Stabilisation}

\author[a]{S. AbdusSalam,}
\author[b]{S. Abel,}
\author[c,d]{M. Cicoli,}
\author[e]{F. Quevedo}
\author[f]{and P. Shukla}

\affiliation[a]{Department of Physics, Shahid Beheshti University, Tehran 19839, Islamic Republic of Iran}
\vspace{2mm}
\affiliation[b]{Institute for Particle Physics Phenomenology, Durham University, South Road, Durham, UK}
\vspace{2mm}
\affiliation[c]{Dipartimento di Fisica e Astronomia, Universit\`a di Bologna, via Irnerio 46, 40126 Bologna, Italy}
\vspace{2mm}
\affiliation[d]{\small INFN, Sezione di Bologna, viale Berti Pichat 6/2, 40127 Bologna, Italy}
\vspace{2mm}
\affiliation[e]{\small DAMTP, University of Cambridge, Wilberforce Road,  Cambridge, CB3 0WA, UK}
\affiliation[f]{\small ICTP, Strada Costiera 11, Trieste 34151, Italy}

\abstract{Achieving full moduli stabilisation in type IIB string compactifications for generic Calabi-Yau threefolds with hundreds of K\"ahler moduli is notoriously hard. This is due not just to the very fast increase of the computational complexity with the number of moduli, but also to the fact that the scalar potential depends in general on the supergravity variables only implicitly. In fact, the supergravity chiral coordinates are 4-cycle volume moduli but the K\"ahler potential is an explicit function of the 2-cycle moduli and inverting between these two variables is in general impossible. In this paper we propose a general method to fix all type IIB K\"ahler moduli in a systematic way by working directly in terms of 2-cycle moduli: on one side we present a `master formula' for the scalar potential which can depend on an arbitrary number of K\"ahler moduli, while on the other we perform a computer-based search for critical points, introducing a hybrid Genetic/Clustering/Amoeba algorithm and other computational techniques. This allows us to reproduce several known minima, but also to discover new examples of both KKLT and LVS models, together with novel classes of LVS minima without diagonal del Pezzo divisors and hybrid vacua which share some features with KKLT and other with LVS solutions.}


\begin{document}

\makeatletter
\let\old@fpheader\@fpheader
\renewcommand{\@fpheader}{\old@fpheader\hfill
%
IPPP/20/18
\hfill }
\makeatother

\maketitle

\bigskip

\section{Introduction}
\label{sec_intro}

Stabilising the moduli fields that determine the size and shape of the extra dimensions has been one of the most important challenges for string compactifications for decades. Flux compactifications of type IIB string theory are probably the most explored since 3-form fluxes stabilise the complex structure moduli $U^\alpha$ (counted via $\alpha=1, \cdots, h^{1,2}$) and the dilaton $S$, producing a huge landscape of solutions. Conversely the K\"ahler moduli $T_i$ (counted by $i=1,\cdots, h^{1,1}$) can be fixed only after perturbative and non-perturbative corrections to the K\"ahler potential and superpotential are included. This last stage of the stabilisation is under less control, due to the difficulty of computing quantum corrections, and of writing the scalar potential explicitly in terms of the correct 4D supergravity chiral coordinates.  

This issue becomes evident when we recall that the imaginary parts of the $T_i$ fields include the 4-cycle volume moduli $\tau_i$ which also give the gauge couplings of the gauge theories living on D7-branes wrapped around internal 4-cycles. Thus the $T$-moduli appear directly in the non-perturbative superpotential. On the other hand the tree-level K\"ahler potential depends directly on the overall Einstein-frame volume:
\be
\vo = \frac16 \,k_{ijk} \, t^i\, t^j \, t^k\,,
\ee
where $k_{ijk}$ are the triple intersection numbers of the underlying Calabi-Yau (CY) threefold and $t^i$ are 2-cycle volumes. The 4-cycle moduli $\tau_i$ are determined in terms of their dual $t^i$ as:
\be
\tau_i = \frac{\partial{\cal V}}{\partial t^i} = \frac12 \, k_{ijk} t^j t^k\,.
\label{eqtau}
\ee
In order to write the full effective action in terms of the $T$-fields, (\ref{eqtau}) needs to be inverted to express $t^i$ as a function of $\tau_j$ which can only be done for simple cases. However a generic CY compactification features a large number of K\"ahler moduli, typically of order $h^{1,1}=100 -1000$ (see for instance \cite{Candelas:1987kf, Candelas:1989hd, Constantin:2016xlj} for classifications on complete intersection CY manifolds and \cite{Kreuzer:2002uu, Kreuzer:2000xy} for CY manifolds as hypersurfaces in toric ambient varieties).\footnote{See also \cite{Cicoli:2012vw, Altman:2014bfa} for partial classifications. Furthermore, it has recently been found that CY manifolds with a comparably large number of moduli have interesting and distinctive properties \cite{Demirtas:2018akl, Braun:2017nhi}.} 

As well as the computational complexity of finding the minimum of a potential with several variables, the fact that in general the scalar potential depends only implicitly on the 4-cycle moduli creates a hard technical obstacle to finding explicit vacua for CY compactifications with large $h^{1,1}$. For this reason the vast majority of  work in the literature has so far focused only on simple examples such as the original KKLT model \cite{Kachru:2003aw} for $h^{1,1}=1$, and vanilla LVS vacua for Swiss cheese and K3-fibred compactifications with $h^{1,1}=2,3$ and diagonal del Pezzo (dP) divisors, where (\ref{eqtau}) can be inverted exactly \cite{Balasubramanian:2005zx, Conlon:2005ki, Cicoli:2007xp}. 

In this paper we propose a new approach to type IIB K\"ahler moduli stabilisation which allows one to overcome these technical issues. Our key idea is to work directly in terms of the 2-cycle volume moduli which appear explicitly in the scalar potential. In combination with a computer-based search this can in principle discover the critical points for an arbitrary number of K\"ahler moduli. Indeed we shall present a `master formula' for the scalar potential generated by $\alpha'$ corrections to the K\"ahler potential and non-perturbative contributions to the superpotential which is valid for arbitrary numbers of 2-cycle moduli. Our subsequent numerical analysis then exploits both Lipschitz optimisation and a hybrid Genetic/Clustering/Amoeba algorithm. 

For convenience, we will illustrate the efficiency of our general method by focusing on CY examples that still have a relatively small number of moduli, but our analysis should be considered as a first step towards tackling more general cases with much larger $h^{1,1}$. In fact, even though in this paper we focus on $h^{1,1}\leq 3$, our method is already able to reveal the existence of entirely new classes of vacua. More precisely, we first show how our `master formula' for the scalar potential in terms of 2-cycle volume moduli combined with our numerical techniques can reproduce several known models, such as standard AdS KKLT vacua \cite{Kachru:2003aw}, dS KKLT solutions with $\alpha'$ uplift \cite{Balasubramanian:2004uy, Westphal:2006tn, Rummel:2011cd, Ben-Dayan:2013fva} and both AdS and dS LVS minima \cite{Balasubramanian:2005zx, Conlon:2005ki, Cicoli:2007xp}. But we then go on to show that our method can also find new examples of both KKLT and LVS models and, more interestingly, it can uncover entirely new classes of LVS minima without diagonal dP divisors, and hybrid vacua that share features of both the KKLT and the LVS solutions.

The developments that we make towards establishing a new systematic approach to K\"ahler moduli stabilisation can be summarised as follows:
\ben
\item We present a `master formula' for the scalar potential as a function of the 2-cycle moduli for an arbitrary number of K\"ahler moduli. The scalar potential is generated by generic single-instanton non-perturbative contributions to the superpotential and the leading $\mathcal{O}(\alpha'^3)$ correction to the K\"ahler potential \cite{Becker:2002nn}. In this first-step approach to K\"ahler moduli stabilisation we neglect string loop corrections \cite{Berg:2005ja,Cicoli:2007xp,Berg:2007wt} and $F^4$ $\mathcal{O}(\alpha'^3)$ contributions \cite{Ciupke:2015msa} which depend explicitly on 2-cycle moduli, providing further motivation for the idea of working directly in terms of the $t$-fields. Given that these corrections are suppressed with respect to the leading $\mathcal{O}(\alpha'^3)$ contribution by either direct powers of $g_s\ll 1$ or inverse powers of the internal volume $\vo \gg 1$, it is consistent to neglect them, although we will include them  in future work which will provide a more comprehensive analysis. For this work we should note that string loops have been used to fix the moduli in \cite{Cicoli:2008va, Cicoli:2008gp, Cicoli:2011qg, Cicoli:2016xae, Cicoli:2017axo, Burgess:2010bz} for simple K3-fibred LVS models with 1 or 2 diagonal dPs, while ref. \cite{Cicoli:2016chb} showed that $F^4$ $\mathcal{O}(\alpha'^3)$ effects can fix all the K\"ahler moduli of any CY with arbitrary large $h^{1,1}$ and at least a single dP divisor. In addition the stabilisation of an arbitrarily large number of K\"ahler moduli in \cite{Cicoli:2016chb} has been achieved by minimising analytically with respect to 2-cycle moduli.

\item We consider the large database of CY threefolds constructed by Kreuzer and Skarke as hypersurfaces in toric ambient varieties \cite{Kreuzer:2000xy}, and we identify the models with $h^{1,1}=1,2,3$ that can be treated with KKLT and LVS techniques (recall LVS needs at least two 4-cycles, a `big' and a `small' divisor). In addition, we find the percentage of models where the relation between 2- and 4-cycle volume moduli cannot be inverted explicitly, so identifying those models that cannot be studied by standard stabilisation techniques,
 where our new method is particularly powerful.

\item We introduce new powerful computational tools for locating and identifying local minima, for example a hybrid Genetic+Clustering+Nelder-Mead algorithm. This technique is of general applicability in systems with many local minima. Due to the computational complexity of identifying local minima we will in practice combine of  such numerical approaches with analytical techniques. 

\item We recover from our `master formula' all the main scenarios that have been proposed so far, for both AdS and dS vacua, including KKLT, $\alpha'$-uplift and LVS, by specifying just 3 quantities: the CY Euler number, the Hodge number $h^{1,1}$ and the number of non-perturbative contributions to the superpotential. 

\item We focus on CY threefolds whose volume does not admit a simple expression in terms of 4-cycle volume moduli, and we find new concrete examples of KKLT vacua for $h^{1,1}=2$ and LVS minima for CY compactifications with $h^{1,1}=3$ and just a single diagonal dP divisor. Moreover we find the first examples in the literature of LVS models for CY compactifications with 3 K\"ahler moduli none of which is a diagonal dP 4-cycle. We also discover novel vacua for CY threefolds with $h^{1,1}=2$ and no diagonal dP 4-cycle. We call these entirely new solutions `hybrid', because the value of the volume at the minimum scales as in KKLT models but the effects used to stabilise the moduli are the same as in LVS models.
\een

This paper is organised as follows. In the Sec. \ref{sec_preliminaries}, after collecting all conventions, we provide the general expression for the scalar potential of the K\"ahler moduli parametrised by the 2-cycle moduli, and we then discuss  the conditions that have to be satisfied to make the effective field theory trustable. In Sec. \ref{Concrete} we make a systematic study of all the models in the Kreuzer-Skarke list with Hodge numbers $h^{1,1}=1,2,3$, classifying those that have a structure admitting LVS vacua and those whose volume form cannot be written explicitly in terms of 4-cycle moduli. In Sec. \ref{NewIIBVacua}  we then show the power of our general method by first reproducing known AdS and dS vacua, and then discovering novel classes of stabilised vacua. Our conclusions are finally presented in Sec. \ref{Conclusions}. We have also collected several technical details in the appendices, starting with App. \ref{AppA} which shows how the scalar potential of several known models can be easily read off from our `master formula'.  The details of the codes used to search for global and also local minima are then explained in Apps. \ref{AppC} and \ref{AppD}, the first describing the Lipschitz optimisation algorithm and the second a hybrid of a genetic algorithm and Clustering and Nelder-Mead algorithms. All of these methods (with their various strong and weak points) are used in combination with analytic calculations to properly identify the local minima. Finally we have included  tables of CY models with $h^{1,1}=2,3$ in Apps. \ref{AppE} and \ref{AppF}.

\section{Type IIB effective theory}
\label{sec_preliminaries}

\subsection{Type IIB preliminaries}

The $F$-term contributions to the ${\cal N}=1$ scalar potential governing the dynamics of low energy effective supergravity are computed from the K\"ahler potential $\mathcal{K}$ and the holomorphic superpotential $W$ via the following well-known relation:
\be
\label{eq:Vtot}
V=e^{\cal K}\Big({\cal K}^{{\cal A}\bar {\cal B}} D_{\cal A} W\, D_{\ov{\cal B}} \ov W-3\, |W|^2\Big)  \,,\,
\ee
where the covariant derivatives are defined with respect to all the chiral variables on which ${\cal K}$ and $W$ generically depend. 

\subsubsection{Fixing the conventions}

The massless states in the 4D effective theory are in one-to-one correspondence with harmonic forms which are either even or odd under the action of an isometric, holomorphic involution $\sigma$ acting on the internal CY threefold, and these generate the equivariant cohomology groups $H^{p,q}_\pm (X)$. Let us fix our conventions and denote the bases of even/odd 2-forms as $(\mu_i, \, \nu_a)$ while 4-forms are denoted $(\tilde{\mu}^i, \, \tilde{\nu}^a)$ where $i=1,..., h^{1,1}_+(X), \, a=1,..., h^{1,1}_-(X)$. Configurations with $h^{1,1}_-(X) \neq 0$ have been studied much less than the simpler $h^{1,1}_-(X) = 0$ case, and explicit constructions of such orientifold odd 2-cycles can be found in \cite{Lust:2006zg,Lust:2006zh,Blumenhagen:2008zz,Cicoli:2012vw,Gao:2013rra,Gao:2013pra}. Also, we denote the 0- and 6-forms as ${\bf 1}$ and $\Phi_6$ respectively. In addition, the bases for the even and odd cohomologies of 3-forms $H^3_\pm(X)$ are denoted respectively as the symplectic pairs $(a_K, b^J)$ and $({\cal A}_\Lambda, {\cal B}^\Delta)$. Using the conventions of \cite{Robbins:2007yv}, let us fix the normalisation in the various cohomology bases as:
\bea
&& \int_X \Phi_6 = 1, \quad \int_X \, \mu_i \wedge \tilde{\mu}^j= \delta_i^{\, \, \,j} , \quad \int_X \, \nu_a \wedge \tilde{\nu}^b = {\delta}_a^{\, \, \,b}, \quad \int_X \, \mu_i \wedge \mu_j \wedge \mu_k = k_{ijk},  \nn \\
&& \int_X \, \mu_i \wedge \nu_a \wedge \nu_b = \hat{k}_{i a b}, \qquad \int_X a_K \wedge b^J = \delta_K{}^J, \qquad \int_X {\cal A}_\Lambda \wedge {\cal B}^\Delta = \delta_\Lambda^\Delta. 
\label{eq:intersection}
\eea
For the orientifold choice with O3/O7-planes, $K=1, ..., h^{2,1}_+$ and $\Lambda=0, ..., h^{2,1}_-$, while for O5/O9-planes, one has $K=0, ..., h^{2,1}_+$ and $\Lambda= 1, ..., h^{2,1}_-$. 

The various fields can be expanded in appropriate bases of the equivariant cohomologies. For example, the K\"ahler form $J$, the 2-forms $B_2$, $C_2$ and the RR 4-form $C_4$ can be expanded as \cite{Grimm:2004uq}:
\bea
J &=& t^i\, \mu_i\,, \qquad  B_2= b^a\, \nu_a\,, \qquad C_2 =c^a\, \nu_a\,,  \nn \\
C_4 &=& \rho_i \, \tilde\mu^i + D_2^i \wedge \mu_i + V^K \wedge a_K + U_K\wedge b^K \,,
\label{eq:fieldExpansions}
\eea
where, as mentioned before, $t^i$ denotes 2-cycle volume moduli, while $b^a$, $c^a$ and $\rho_i$ are various axions. Furthermore ($V^K$, $U_K$) forms a dual pair of space-time 1-forms and $D_2^i$ is a space-time 2-form dual to the scalar field $\rho_i$. Also, since $\sigma^*$ reflects the holomorphic 3-form $\Omega_3$, we have $h^{2,1}_-(X)$ complex structure moduli $U^\alpha$ appearing as complex scalars. Moreover, the involutively-odd holomorphic 3-form $\Omega_3$ generically depends on the complex structure moduli and can be written in terms of the period vectors as: 
\be
\label{eq:Omega3}
\Omega_3\, \equiv  {\cal X}^\Lambda \, {\cal A}_\Lambda - \, {\cal F}_{\Lambda} \, {\cal B}^\Lambda \,,
\ee
where ${\cal F} = ({\cal X}^0)^2 \, \, f({U^\alpha})$ is a generic pre-potential, with $U^\alpha =\frac{\delta^\alpha_\Lambda \, {\cal X}^\Lambda}{{\cal X}^0}$ and with $f({U^\alpha})$ being some function dependent on the complex structure moduli \cite{Hosono:1994av}. Apart from the complex structure moduli, the dynamics of the ${\cal N} = 1$ type IIB 4D effective theory can be described using the following additional chiral variables ($S, G^a, T_i$) defined as in \cite{Benmachiche:2006df}:
\bea
S &=& C_0 + \, {\rm i} \, e^{-\phi} = C_0 + {\rm i}\, s\, , \qquad G^a= c^a + S \, b^a \,, \nn \\
T_i &=& \left(\rho_i +\, \hat{{k}}_{i a b} c^a b^b + \frac12 \, C_0 \, \hat{{k}}_{i a b} b^a \, b^b \right) -{\rm i}\, \left(\tau_i - \frac{s}{2} \, \hat{{k}}_{i a b} \, b^a \, b^b \right)\,, 
\label{eq:N=1_coords}
\eea 
where $\tau_i = \frac12 \, {k}_{ijk} t^j t^k$ is an Einstein frame 4-cycle volume. In addition we will introduce the short-hand notation $k^{ij} \equiv \left(k_{ijk} \, t^k\right)^{-1}$.

At the perturbative level, the K\"ahler potential receives two kinds of corrections: $\alpha'$ and $g_s$ corrections. Using appropriate chiral variables, a generic form for the K\"ahler potential incorporating the leading $\mathcal{O}(\alpha'^3)$ correction can be written as the sum of two terms motivated by their underlying ${\cal N}=2$ special K\"ahler and quaternionic structure:
\be
\label{eq:K}
{\cal K} = K_{\rm cs} + K\,,
\ee
where:
\be
\label{eq:defK}
K_{\rm cs} = -\ln\left({\rm i}\int_X \Omega_3\wedge{\bar\Omega_3}\right) \qquad\text{and}\qquad K = - \ln\left(-{\rm i}(S-\ov{S})\right) -2\ln{\cal Y}\,. 
\ee
Here ${\cal Y}$ denotes the $\alpha'$ corrected CY volume \cite{Becker:2002nn}:
\be
{\cal Y} = \vo + \frac{\xi}{2}\, \left(\frac{S-\ov{S}}{2\,{\rm i}}\right)^{3/2} = \vo +  \frac{\xi}{2\, g_s^{3/2}}\,,
\label{eq:defY}
\ee
where $\vo$ is the tree-level CY volume $\vo = \frac16 \,{k_{ijk} \, t^i\, t^j \, t^k}$ in Einstein frame and $\xi$ is proportional to the CY Euler characteristics $\chi$: $\xi = -\frac{\zeta(3)\, \chi(X)}{2\,(2\pi)^3}$ (for reference $\zeta(3) \simeq 1.2$). Further $\alpha'$ and $g_s$ corrections have been estimated throughout the years, turning out to be either subdominant or reabsorbable by field redefinitions. Finding all the possible $\alpha'$ corrections is an open question. For a recent discussion of these corrections see for instance \cite{Cicoli:2018kdo}. 

The block diagonal nature of the total K\"ahler metric (and its inverse) admits the following splitting of contributions:
\be
\label{eq:V_gen}
e^{- {\cal K}} \, V = {\cal K}^{{\cal A} \ov {\cal B}} \, (D_{\cal A} W) \, (D_{\ov {\cal B}} \ov{W}) -3 |W|^2 \equiv V_{\rm cs} + V_{\rm k}\,,
\ee
where:
\be
\label{eq:VcsVk}
V_{\rm cs} =  K_{\rm cs}^{\alpha \ov {\beta}} \, (D_\alpha W) \, (D_{\ov {\beta}} \ov{W}) \qquad \text{and}\qquad V_{\rm k} =  K^{{A} \ov {B}} \, (D_{A} W) \, (D_{\ov {B}} \ov{W}) -3 |W|^2\,.
\ee
Recall that the indices $(\alpha,\beta)$ correspond to the complex structure moduli $U^\alpha$ while the indices $(A,B)$ run over the remaining chiral variables $\{S, G^a, T_i\}$. For our purposes, we choose the orientifold involution such that the odd $(1,1)$-cohomology sector is trivial, and so there will be no odd moduli present in the current analysis.

\subsubsection{Inverse K\"ahler metric and useful identities}

The derivatives of the K\"ahler moduli dependent piece of the K\"ahler potential ($K$) in (\ref{eq:defK}) can be generically expressed as (with $\hat\xi\equiv \xi/ g_s^{3/2}$):
\be
\label{eq:derK}
K_S = \frac{{\rm i}}{2 \,s }\left(1 + \frac{3\, \hat\xi}{2\,{\cal Y}}\right) = - K_{\ov{S}}, \qquad K_{T_i} = -\frac{{\rm i} \, t^i}{2\, {\cal Y}} = - K_{\ov{T}_i}\,.
\ee
Using these derivatives, the various K\"ahler metric components are found to be:
\be
\label{eq:simpK}
K_{S \ov{S}} = \frac{1}{4\,s^2}\, \left(1 - \frac{3\, \hat{\xi}}{4\, {\cal Y}}+ \frac{9\, \hat{\xi}^2}{8\, {\cal Y}^2} \right), \quad K_{T_i \, \ov{S}} = -\frac{3\, \hat\xi\, t^i}{16\, s\, {\cal Y}^2}= K_{S\,\ov{T}_i}, \quad K_{T_i \, \ov{T}_j} = \frac{9\, {\cal G}^{ij}}{4\, {\cal Y}^2},
\ee
where, using our shorthand notation, the $\alpha'$-corrected moduli space metric and its inverse, ${\cal G}$ and ${\cal G}^{-1}$, are given by:
\be
\label{eq:genMetrices}
\frac{{\cal G}_{ij}}{36} = \frac{\tau_i \,\tau_j}{{\cal Y}\, (6\vo - 2\, {\cal Y})} -\frac{k_{ijk} t^k}{4\, {\cal Y}} \qquad\text{and}\qquad 36\,{\cal G}^{ij} = 2 \, t^i \, t^j -4\, {\cal Y} \, k^{ij}. 
\ee 
Hence the inverse K\"ahler metric components are found to be \cite{Bobkov:2004cy}:
\be
K^{{S} \ov{S}} =   \gamma_1, \qquad K^{T_i \, \ov{S}} =  \gamma_2\, \tau_i = K^{{S}\,\ov{T}_i}, \qquad
K^{T_i \, \ov{T}_j} = \frac49\, {\cal Y}^2\, {\cal G}_{ij} + \frac{\gamma_2^2}{\gamma_1}\, \tau_i\,\tau_j\,, 
\label{eq:InvK}
\ee
where $\gamma_1$ and $\gamma_2$ are given by:
\be
\label{eq:gamma123}
\gamma_1 = \frac{s^2\,(4 \,{\cal V}-\hat{\xi})}{(\vo-\hat\xi)}\,, \quad \qquad \gamma_2 = \frac{3\, s\,\hat{\xi}}{\,({\cal V}-\hat{\xi})}\,\,. 
\ee
Here let us note that in the absence of $\alpha'$ corrections, i.e. setting $\hat\xi= 0$, we have $\gamma_1 = 4\, s^2$ and $\gamma_2 = 0$, and 
the inverse metric components in (\ref{eq:InvK}) reduce to the standard results of \cite{Grimm:2004uq}. Considering the explicit components of the inverse K\"ahler metric, we find the following useful simplified relations:
\bea
K_S\,K^{S\ov{S}} &=& \frac{{\rm i}\,s\,(4 \,\vo-\hat\xi) (\vo + 2\,\hat\xi)}{({\cal V}-\hat{\xi})(2\vo + \hat{\xi })} = -\, K^{S \ov{S}} \, K_{\ov{S}},  \nn \\
K_S \, K^{{S} \ov{T}_i}  &=& \frac{3 \, {\rm i} \, \hat\xi\, \tau_i \, (\vo + 2 \, \hat{\xi })}{({\cal V}-\hat{\xi }) (2\vo+\hat{\xi})} = -\, K^{{T_i} \ov{S}} \, K_{\ov{S}},   \nn \\
K_{T_i} \, K^{T_i\,\ov{S}} &=& -\frac{9 \, {\rm i} \, s\, \hat{\xi} \,{\cal V}}{({\cal V}-\hat{\xi}) (2\vo+\hat{\xi })} 
= -\, K^{S\,\ov{T}_i} \, K_{\ov{T}_i},  \nn \\
K_{T_i} \, K^{{T_i} \ov{T}_j}  &=& -\frac{i \, \tau_j \,(4 \,{\cal V}^2+\vo \,\hat\xi +4 \, \hat\xi^2)}{({\cal V}-\hat{\xi }) (2\vo+\hat{\xi })} = -\, K^{{T_j} \ov{T}_i} \, K_{\ov{T}_i}\,, 
\label{eq:identities1}
\eea
together with:
\bea
K_S \, K^{S \ov{S}} \, K_{\ov{S}} &=& \frac{(4 \,{\cal V}-\hat\xi) (\vo+2 \,\hat{\xi })^2}{({\cal V}-\hat{\xi}) (2\vo+\hat{\xi })^2} \, , \nn \\
K_S \, K^{{S} \ov{T}_i} \, K_{\ov{T}_i} &=& -\frac{9 \,\hat{\xi} \,{\cal V} (\vo+ 2 \, \hat{\xi })}{({\cal V}-\hat{\xi }) (2\vo + \hat\xi)^2} = K_{T_i} \, K^{{T_i} \ov{S}} \, K_{\ov{S}} \,, \nn \\
K_{T_i} \, K^{{T_i} \ov{T}_j} \, K_{\ov{T}_j} &=& \frac{3 \,{\cal V} (4 \,{\cal V}^2+\vo\,\hat{\xi}+4 \, \hat{\xi}^2)}{({\cal V}-\hat{\xi}) (2\vo + \hat{\xi})^2} \,. 
\label{eq:identities2}
\eea
These identities will be used extensively in the derivation of the master formula for the scalar potential. As a check, when $\alpha'$ corrections are turned off, i.e. when $\hat\xi=0$, these useful identities reduce to the following well-known tree-level results:
\bea
K_S \, K^{S\ov{S}} &=& 2\,{\rm i}\,s = -\, K^{S \ov{S}} \, K_{\ov{S}}, \quad K_S \, K^{{S} \ov{T}_i} = 0 = K^{{T_i} \ov{S}} \, K_{\ov{T}_i}, \nn \\
K_{T_i} \, K^{T_i\,\ov{S}} &=& 0 = K^{S\,\ov{T}_i} \, K_{\ov{T}_i}, \quad K_{T_i} \, K^{{T_i} \ov{T}_j} = - 2\, {\rm i}\, \tau_j = -\, K^{{T_j} \ov{T}_i} \, K_{\ov{T}_i}, \nn \\
K_S \, K^{S \ov{S}} \, K_{\ov{S}} &=& 1, \quad K_{S} \, K^{{S} \ov{T}_i} \, K_{\ov{T}_i} = 0, \quad K_{T_i} \, K^{{T_i} \ov{T}_j} \, K_{\ov{T}_j} = 3\,.
\eea

\subsection{A master formula for the scalar potential}

For a generic superpotential which depends on all closed string chiral variables, namely $S$, $T_i$ and $U^\alpha$, the $F$-term scalar potential (\ref{eq:Vtot}) can be rewritten as:
\bea
e^{- {\cal K}} \, V \,&=& \,  K_{\rm cs}^{\alpha \ov {\beta}} \, (D_\alpha W) \, (D_{\ov{\beta}} \ov{W}) + K^{S \ov{S}} \, (D_S W) \, (D_{\ov{S}} \ov{W}) +K^{S \ov{T}_i} \, K_{\ov{T}_i}\, (D_S W)  \, \ov{W} \nn \\
&+&\, K_{T_i} \,K^{{T_i} \ov{S}}  \, (D_{\ov{S}} \ov{W}) \,  W +K^{S \ov{T}_i} \, (D_S W) \, \ov W_{\ov{T}_i}  +K^{{T_i} \ov{S}} \, W_{T_i} \, (D_{\ov{S}} \ov{W}) \nn \\
&+& K_{T_i} K^{{T_i} \ov{T}_j}  W \ov{W}_{\ov{T}_j} + W_{T_i} K^{{T_i} \ov{T}_j} K_{\ov{T}_j} \ov{W} + W_{T_i} K^{{T_i} \ov{T}_j} \ov{W}_{\ov{T}_j} \nn \\
&+& \left(K_{T_i} \, K^{{T_i} \ov{T}_j} \, K_{\ov{T}_j} -3 \right)|W|^2 \,. 
\label{eq:Vgen0}
\eea
In standard flux compactifications with $F_3$ and $H_3$ fluxes, the tree-level superpotential depends only on the complex structure moduli and the axio-dilaton, i.e. $W = W_0(U^\alpha, S)$. This flux-dependent superpotential can fix all complex structure moduli and the axio-dilaton supersymmetrically at leading order by enforcing:
\be
D_\alpha W_0 = 0 =  D_{\ov{\alpha}} \, \ov{W}_0, \qquad\text{and}\qquad D_S W_0 \, = 0 = D_{\ov{S}} \, \ov{W}_0\,.
\label{UStab}
\ee
The K\"ahler moduli can appear in $W$ only via non-perturbative effects. In what follows, we shall assume $n$ non-perturbative contributions to $W$ which can be generated by either rigid divisors, such as shrinkable dP 4-cycles, or non-rigid divisors with non-zero magnetic fluxes \cite{Bianchi:2011qh, Bianchi:2012pn, Louis:2012nb}. The corresponding non-perturbative superpotential is then:\footnote{The exponents $(- {\rm i}\, a_i\, T_i)$ in (\ref{eq:Wnp-n}) follow from the definition of the chiral variables in (\ref{eq:N=1_coords}) which have been chosen to make explicit the T-duality transformations between type IIA and type IIB \cite{Shukla:2019wfo}.}
\be
W= W_0 + \sum_{i = 1}^n \, A_i\, e^{- {\rm i}\, a_i\, T_i}\,.
\label{eq:Wnp-n}
\ee
Note that in (\ref{eq:Wnp-n}) there is no sum in the exponents $(- {\rm i}\, a_i\, T_i)$, and summations are to be understood only when upper indices are contracted with lower indices; otherwise we will write an explicit sum as in (\ref{eq:Wnp-n}). We will suppose that out of $h^{1,1}$ K\"ahler moduli, only the first $n$ appear in $W$, i.e. $i=1,...,n\leq h^{1,1}$. 

Assuming that the $S$ and $U$-moduli are stabilised as in (\ref{UStab}) and considering a superpotential given by (\ref{eq:Wnp-n}), the scalar potential (\ref{eq:Vgen0}) reduces to:
\be
V =  e^{{\cal K}} \, \left[ K_{T_i} K^{{T_i} \ov{T}_j}  W \ov{W}_{\ov{T}_j} + W_{T_i} K^{{T_i} \ov{T}_j} \left(\ov{W}_{\ov{T}_j}+ K_{\ov{T}_j} \ov{W} \right) + \left(K_{T_i} \, K^{{T_i} \ov{T}_j} \,  K_{\ov{T}_j} -3 \right)|W|^2 \right]\,. \nn
\ee
Moreover, using the identities in eqs.~(\ref{eq:identities1})-(\ref{eq:identities2}) which include $\mathcal{O}(\alpha'^3)$ corrections to the K\"ahler potential, this scalar potential can be written as the sum of three terms:
\be
V = V_{\mathcal{O}(\alpha'^3)} + V_{\rm np1} + V_{\rm np2} \,,
\label{eq:Vgen-nGen}
\ee
where (introducing phases into the parameters as $W_0=|W_0|\, e^{{\rm i} \, \theta_0}$ and $A_i = |A_i|\, e^{{\rm i}\, \phi_i}$):
\bea
\label{MasterF}
V_{\mathcal{O}(\alpha'^3)} &=&  e^{{\cal K}} \, \frac{3 \, \hat\xi (\vo^2 + 7\,\vo\, \hat\xi +\hat\xi^2)}{({\cal V}-\hat{\xi }) (2\vo + \hat{\xi })^2}\, \,|W_0|^2\,, \\
V_{\rm np1} &=& e^{{\cal K}}\, \sum_{i =1}^n \, 2 \, |W_0| \, |A_i|\, e^{- a_i \tau_i}\, \cos(a_i\, \rho_i + \theta_0 - \phi_i) \nn \\
&& \times ~\biggl[\frac{(4 \vo^2 + \vo \, \hat\xi+ 4\, \hat\xi^2)}{ (\vo - \hat\xi) (2\vo + \hat\xi)}\, (a_i\, \tau_i) +\frac{3 \, \hat\xi (\vo^2 + 7\,\vo\, \hat\xi + \hat\xi^2 )}{(\vo-\hat\xi) (2\vo + \hat\xi)^2}\biggr]\,, \nn \\
V_{\rm np2} &=& e^{{\cal K}}\, \sum_{i=1}^n\,  \sum_{j=1}^n \, |A_i|\, |A_j| \, e^{-\, (a_i \tau_i + a_j \tau_j)} \, \cos(a_i\, \rho_i - a_j \, \rho_j -\phi_i + \phi_i) \,\nn \\
&& \times \biggl[ -4 \left({\cal V}+\frac{\hat\xi}{2}\right) \, (k_{ijk}\,t^k) \, a_i\, a_j\, + \frac{4{\cal V} - \hat{\xi}}{(\vo - \hat\xi)} \left(a_i\, \tau_i) \, (a_j\,\tau_j \right) \nn \\
&& + \,\frac{(4\vo^2  + \vo \, \hat{\xi} + 4\, \hat{\xi}^2)}{(\vo - \hat{\xi}) (2\vo + \hat{\xi})}\, (a_i\, \tau_i +a_j\, \tau_j) +\frac{3 \, \hat{\xi} (\vo^2 + 7\,\vo\, \hat{\xi}  +\hat\xi^2)}{({\cal V}-\hat{\xi }) (2\vo + \hat{\xi })^2}\biggr]~. \nn
\eea
Notice that $V_{\mathcal{O}(\alpha'^3)}$ reproduces the known $\mathcal{O}(\alpha'^3)$ contribution to the potential first derived in \cite{Becker:2002nn}. This term vanishes for $\hat\xi = 0$, reproducing the standard no-scale structure in the absence of a $T$-dependent non-perturbative $W$. On the other hand, for very large volume $\vo \gg \hat\xi$, this term takes the standard form which plays a crucial r\^ole in LVS models \cite{Balasubramanian:2005zx}:
\be
V_{\mathcal{O}(\alpha'^3)} \simeq \frac{e^{K_{\rm cs}}}{2\,s\, \vo^2}\, \times \frac{3\,\hat\xi\, |W_0|^2}{4\, {\cal V}}\,.
\ee
Let us also stress that $V_{\mathcal{O}(\alpha'^3)}$ depends only on the overall volume $\vo$, while $V_{\rm np1}$ depends on $\vo$ and the 4-cycle moduli $\tau_i$ (with the additional dependence on the axions $\rho_i$). Hence these two contributions to $V$ could be minimised by taking derivatives with respect to $\vo$ and $(h^{1,1}-1)$ 4-cycle moduli. However $V_{\rm np2}$ depends on the quantity $k_{ijk}\,t^k$ which in general cannot be inverted to be expressed as an explicit function of the $\tau_i$'s. Thus our master formula for the scalar potential shows that moduli stabilisation is more naturally performed in terms of the 2-cycle moduli $t^i$. As discussed in the introduction, we will see that this strategy allows the study of a much wider set of cases, leading to new interesting moduli stabilisation schemes.

Moreover (\ref{MasterF}) determines the complete form of $V$ simply by specifying topological quantities such as the intersection numbers and the CY Euler number which controls $\mathcal{O}(\alpha'^3)$ corrections, and the number $n$ of non-perturbative contributions to $W$. Before proceeding to find new vacua, in Tab. \ref{tab_known-models} we show how our master formula can elegantly reproduce known moduli stabilisation models parametrised by different choices of $h^{1,1}$, $n$ and $\hat\xi$ (see App. \ref{AppA} for an explicit derivation of these potentials from our master formula). 

\begin{center}
\begin{tabular}{ |c||c|c|c| } 
\hline 
Model \qquad & $h^{1,1}$ \qquad & $n$ \qquad & $\hat\xi$ \qquad \\ 
\hline
1-modulus KKLT \cite{Kachru:2003aw} \quad & \quad $h^{1,1} = 1$ \qquad & $n = 1$  & \quad $\hat\xi =0$ \qquad \\
1-modulus KKLT with $\alpha'$-uplift \cite{Westphal:2006tn, Rummel:2011cd, Ben-Dayan:2013fva} \qquad & \quad $h^{1,1}=1$ \quad & $n = 1$ & \quad $\hat\xi>0$ \quad \\
2-moduli KKLT \cite{Denef:2004dm, BlancoPillado:2006he} \qquad & \quad $h^{1,1}=2$ \quad & $n = 2$ & \quad $\hat\xi=0$ \quad \\
2-moduli KKLT with $\alpha'$-uplift \cite{Louis:2012nb} \quad & \quad $h^{1,1}=2$ \quad & $n = 2$ & \quad $\hat\xi>0$ \quad \\
2-moduli Swiss cheese LVS \cite{Balasubramanian:2005zx,Cicoli:2012vw, Cicoli:2013mpa, Cicoli:2013cha} \quad & \quad $h^{1,1} = 2$ \quad & $n = 1$ & \quad $\hat{\xi} > 0$ \quad \\
3-moduli Swiss cheese LVS \cite{Conlon:2005jm, Cicoli:2017shd} \qquad & \quad $h^{1,1}=3$ \quad &  $n = 2$ & \quad $\hat\xi>0$ \quad \\
3-moduli fibred LVS \cite{Cicoli:2008va} \qquad & \quad $h^{1,1}=3$ \quad &  $n = 2$ & \quad $\hat\xi>0$ \quad \\
\hline
\end{tabular}
\captionof{table}{Various classes of known models whose scalar potential can be easily read-off from our master formula (\ref{MasterF}).}
\label{tab_known-models}
\end{center}

Furthermore, our master formula features an explicit dependence on all phases and axion fields. In this paper we shall fix the axions analytically and scan numerically for minima along the directions of the 2-cycle moduli. However (\ref{MasterF}) allows for a more general numerical analysis of the many axion potential. We leave this for future work. 

\subsection{Validity of the effective field theory}
\label{EFT}

Before using the master formula in (\ref{MasterF}) to find new minima, let us list the conditions that have to be satisfied to trust the validity of the low-energy 4D effective field theory:
\ben
\item \textbf{Stringy corrections}: Stringy effects can be neglected if each 2-cycle $\Sigma_2^{(i)}$, $i=1,...,h^{1,1}$, has a string-frame volume larger than the string scale, i.e. ${\rm Vol}_s\left(\Sigma_2^{(i)}\right) \gg \alpha'$ $\forall i=1,...,h^{1,1}$. Given that string and Einstein frame volumes are related as ${\rm Vol}_s \left(\Sigma_2^{(i)}\right) = g_s^{1/2} {\rm Vol}_{\rm E} \left(\Sigma_2^{(i)}\right)$, and expressing the 2-cycle volumes in units of $\ell_s = 2\pi\sqrt{\alpha'}$ as ${\rm Vol}_{\rm E} \left(\Sigma_2^{(i)}\right) = |t_i|\, \ell_s^2$, the condition to trust the supergravity regime is \cite{Cicoli:2017axo}:
\be
|t_i| \gg \frac{1}{g_s^{1/2}\, (2\pi)^2} \qquad \forall\,i=1,...,h^{1,1}\,.
\label{cond1}
\ee
The 10D tree-level action receives higher derivative corrections at different orders in $\alpha'$ which, at the level of the 4D effective theory, appear as an expansion in inverse powers of the K\"ahler moduli. Hence the condition (\ref{cond1}) guarantees that the $\alpha'$ expansion is well-behaved. In what follows, we shall consider only the leading $\mathcal{O}(\alpha'^3)$ correction to the K\"ahler potential in (\ref{eq:defY}) which generates $V_{\mathcal{O}(\alpha'^3)}$ in (\ref{MasterF}). However this expression can be trusted only if higher $\alpha'$ effects can be neglected which requires:
\be
\frac{\xi}{2\,g_s^{3/2}\,\vo}\ll 1\,.
\label{cond2}
\ee

\item \textbf{String loops}: A crucial requirement to trust moduli stabilisation is that perturbation theory does not break down. Given that string loop corrections to the K\"ahler potential are proportional to the string coupling $g_s$, we need therefore to impose: 
\be
g_s^{-1} = e^{-\phi} = \frac{\left(S-\ov{S}\right)}{2\,{\rm i}} \gg 1\,,
\label{gscond}
\ee
which can be met by an appropriate choice of background fluxes that fix $S$. In our analysis we shall neglect $g_s$ corrections to the effective action. As explained in \cite{Cicoli:2007xp, Burgess:2016owb}, this is justified by the existence of an `extended no-scale structure' so that string loop effects start contributing to the effective action only at $\mathcal{O}(g_s^2 \alpha'^4)$. 

\item \textbf{Non-perturbative effects}: The superpotential (\ref{eq:Wnp-n}) contains only single-instanton contributions while in general multi-instanton effects would also be present. These can be safely neglected if:
\be
a_i\, \tau_i \gg 1\qquad\forall\,i=1,...,n\,.
\label{cond3}
\ee

\item \textbf{4D supergravity regime}: The low-energy supergravity theory admits a valid 4D description only if the Kaluza-Klein (KK) modes are heavy. In a string compactification, there can actually be several KK scales $M^{(i)}_{\rm KK}$ associated with either bulk modes or open string excitations on D7-branes wrapped around 4-cycles. We therefore require the following hierarchy of mass scales:
\be
m_{3/2}, m_{\rm mod} \ll M^{(i)}_{\rm KK} \lesssim M_s \ll M_p \qquad\forall\,i\,,
\ee
where $m_{\rm mod}$ denotes generic moduli masses, $m_{3/2}$ is the gravitino mass, $M_s$ is the string scale and $M_p$ is the reduced Planck mass given by (see \cite{Burgess:2010bz} for the proper normalisation factor $\kappa = g_s\,e^{K_{\rm cs}}/(8\pi)$ in 4D Einstein frame):
\bea
m_{3/2} &=& e^{\mathcal{K}/2}\,|W| \simeq \sqrt{\kappa}\, \frac{|W_0|}{\vo}\,M_p\,, \qquad M^{(i)}_{\rm KK} =\frac{\sqrt{\pi}}{\sqrt{\vo}\,\tau_i^{1/4}}\,M_p\,, \nn \\
M_s &\equiv& 1/\sqrt{\alpha'} = \frac{g_s^{1/4}\sqrt{\pi}}{\sqrt{\vo}}\,M_p\,, \qquad M_p = \left(8\pi G\right)^{-1/2} = 2.4 \cdot 10^{18}\,{\rm GeV}\,. 
\eea
The condition $M_s \ll M_p$ $\forall\,i$ is guaranteed by (\ref{cond1}) while $M^{(i)}_{\rm KK} \lesssim M_s$ corresponds to $\tau_i \gtrsim g_s^{-1}$ which is always true for `large' 4-cycles while it is marginally satisfied for relatively `small' moduli. The condition $m_{3/2} \ll M^{(i)}_{\rm KK}$ is more severe when $i={\rm bulk}$ with $\tau_{\rm bulk}\simeq \vo^{2/3}$. In this case we have therefore to impose:
\be
m_{3/2} \ll M^{({\rm bulk})}_{\rm KK} \qquad \Leftrightarrow \qquad \sqrt{\frac{\kappa}{\pi}}\, |W_0| \ll \vo^{1/3}\,,
\label{cond4}
\ee
which sets an important upper bound on the vacuum expectation value of the flux-generated superpotential $|W_0|$. 

\item \textbf{Superspace higher-derivative expansion}: Ref. \cite{Cicoli:2013swa} established that the coupling of heavy bulk KK modes to light states scales as $g \sim M^{({\rm bulk})}_{\rm KK} /M_p \sim \vo^{-2/3} \ll  1$. Denoting the auxiliary field of the light fields as $F\sim m_{3/2} M_p$ and the UV cut-off as $\Lambda \sim M^{({\rm bulk})}_{\rm KK}$, the superspace derivative expansion is therefore under control if $g\,F/ \Lambda^2 \sim m_{3/2}/ M^{({\rm bulk})}_{\rm KK} \ll  1$, which is guaranteed to hold if (\ref{cond4}) is satisfied.  
\een

\section{Explicit CY examples}
\label{Concrete}

In this section we will present a classification of all CY threefolds with $1 \leq h^{1,1} \leq 3$ from the Kreuzer-Skarke list where these manifolds have been constructed via toric geometry \cite{Kreuzer:2000xy}. We will perform this analysis with the help of a database \cite{Altman:2014bfa} which provides several topological properties of all CY threefolds with $1 \leq h^{1,1} \leq 6$ arising from triangulations of the polytopes of the Kreuzer-Skarke list. The most relevant data that we will use are the GLSM charges, the Stanley-Reisner (SR), the intersection tensor, the Mori cone and Euler characteristics. Knowing the GLSM charges along with the SR ideal will enable us to analyse the divisor topologies using the cohomCalg package \cite{Blumenhagen:2010pv, Blumenhagen:2011xn}. We will then perform a choice of divisor basis which takes the overall volume $\vo$ in its simplest possible form and makes some important features for moduli stabilisation manifest (like the presence of diagonal dP divisors for LVS constructions). This will also allow us to divide the models into those where the volume admits a simple form in terms of 4-cycle volume moduli, and those where it does not. 

This classification will then be used in Sec. \ref{KKLTstab} and \ref{NewVacua} to show how our master formula can be used to stabilise the K\"ahler moduli in generic situations where the volume can be expressed in a simple way only as a function of 2-cycle moduli. In fact, we will show that converting 2- into 4-cycle volume moduli can be hard even for simple examples where only a few intersection numbers are non-zero. In what follows we denote the various models as $M_{i,j}$, where $i$ indicates the value of $h^{1,1}$ while $j$ labels a given CY threefold at fixed $h^{1,1}$.

\subsection{$h^{1,1}=1$}

In the presence of a single K\"ahler modulus the conversion from $t$ to $\tau$ is trivially possible. In this case the Kreuzer-Skarke database features 5 distinct CY threefolds whose details relevant for moduli stabilisation are summarised in Tab. \ref{tab_cydata-h11eq1}.

\begin{center}
\begin{tabular}{|c||c|c|c|c|c|c||} 
\hline
Model & $\chi$ & $k_{111}$  & $\tau_1$ & $\vo$ & K\"ahler cone \\
\hline
$M_{1,1}$ &-40  & 1 & $\tfrac12\, t_1^2$ & $\tfrac16\,t_1^3$ & $t_1 > 0$ \\
$M_{1,2}$ &-200 & 5 & $\tfrac52 \, t_1^2$ & $\tfrac56\, t_1^3$ & $t_1 > 0$ \\
$M_{1,3}$ &-204 & 3 & $\tfrac32 \, t_1^2$ & $\tfrac12\,t_1^3$ & $t_1 > 0$ \\
$M_{1,4}$ &-288 & 1 & $\tfrac12 \,t_1^2$ & $\tfrac16\,t_1^3$ & $t_1 > 0$ \\
$M_{1,5}$ &-296 & 2 & $t_1^2$ & $\tfrac13\,t_1^3$ & $t_1 > 0$ \\
\hline
\end{tabular}
\captionof{table}{Relevant data for CY geometries with $h^{1,1}=1$.}
\label{tab_cydata-h11eq1}
\end{center}

\subsection{$h^{1,1}=2$}

In the Kreuzer-Skarke database there are 36 reflexive polytopes with $h^{1,1}= 2$ leading to 48 triangulations \cite{Altman:2014bfa}. However given that different polytopes can lead to the same triangulations, there are only 39 distinct CY geometries listed in Tab. \ref{tab_cydata-h11eq2} in App. \ref{AppE}. As can be seen from Tab. \ref{tab_cydata-h11eq2}, in all cases 1 intersection number can always be eliminated by an appropriate choice of basis.\footnote{Here we limit the discussion to the coordinate divisors but there may be some non-toric divisor combinations that reduce this number further. However, such cases are likely to be non-smooth and hence they are not suitable for phenomenology.} All these 39 models with $h^{1,1}=2$ can be classified as follows:
\bi
\item 22 CY geometries feature 1 diagonal dP 4-cycle, allowing them to be written in the strong Swiss cheese form $\vo \sim \tau_1^{3/2} - \tau_2^{3/2}$. The negative sign arises from the K\"ahler cone condition $t_1 < 0$ which characterises all LVS models, as can be seen in Tab. \ref{tab_cydata-h11eq2}. 

\item 10 CY threefolds are K3-fibred. In these cases at least 2 intersection numbers can be removed by an appropriate choice of basis. For example, if $D_1$ is the K3 divisor, one can always find another suitable divisor $D_2$ to form a basis where $k_{111}  = k_{112} = 0$, as a consequence of a theorem for K3-fibred CY threefolds \cite{Oguiso:1993, Schulz:2004tt}. Moreover, in some cases (like $\mathbb{C}{\rm P}^4[1,1,2,2,2]$), it is even possible to make $k_{222} = 0$, leaving $k_{122}$ as the only non-zero intersection number and $\vo \sim t_1\, t_2^2 \sim \sqrt{\tau_1}\,\tau_2$. As can be seen from Tab. \ref{tab_cydata-h11eq2}, $M_{2, 33}$ is an example with this simple form of the CY volume. 

\item 7 CY threefolds do not admit a simple volume form in terms of 4-cycle volume moduli. These are the examples which are of interest to us and are highlighted as `hard' in Tab. \ref{tab_cydata-h11eq2}. The simplest example in this class of models is $M_{2,1}$ with volume form:
\be
\vo = \frac12\left(t_1^2\, t_2 + t_1\, t_2^2\right).
\ee
This simple example already illustrates the difficulty to invert the relations between 2- and 4-cycle volume moduli which look like:
\be
\tau_1 = t_1\, t_2 + \frac12\,t_2^2\,, \qquad \tau_2 = t_1\, t_2 + \frac12\,t_1^2\,.
\ee
The conversion from 2- to 4-cycle moduli results in the following 4 sets of solutions:
\be
t_1 = \pm \sqrt{\frac23} \, x_\pm, \qquad  t_2 =  \pm\,\frac{\left(4\, \tau_1 - 3\, \tau_2 + x_\pm^2\right)}{\sqrt{6}\, \tau_2}\,x_\pm\,, 
\label{solutions}
\ee
where:
\be
x_\pm = \sqrt{\tau_2 - 2\, \tau_1 \pm 2 \, \sqrt{\tau_1^2 - \tau_1\, \tau_2 + \tau_2^2}}\,.
\ee
A unique solution is identified by the K\"ahler cone conditions $\{t_1 > 0, \, t_2 > 0\}$ which select in (\ref{solutions}) the $x_+$-dependent solution with positive signs. In order to illustrate our numerical analysis, in what follows we shall consider $M_{2,6}$ and $M_{2,20}$ as representative benchmark examples of this class of `hard' CY models.
\ei

\subsection{$h^{1,1}=3$}
\label{h113}

In the $h^{1,1} = 3$ case, a generic CY geometry features 10 intersection numbers. In the Kreuzer-Skarke database, there are 244 reflexive polytopes for $h^{1,1}= 3$ leading to 569 triangulations \cite{Altman:2014bfa}. However given that different polytopes can lead to the same triangulation, there are only 305 distinct CY geometries which we classify as follows:\footnote{The number of distinct CY examples is actually 306 but 1 is a non-favorable geometry which we do not consider relevant for phenomenology.}
\bi
\item 232 CY geometries have at least 1 divisor, say $D_p$, whose corresponding intersection numbers satisfy the following condition first derived in \cite{Cicoli:2018tcq}:
\be
k_{ppi}\,k_{ppj} = k_{ppp}\,k_{pij}\qquad \forall\,i,j\,.
\label{Dpcondition}
\ee
This condition allows to trade easily one 2-cycle modulus for $\tau_p$ since it guarantees that $\tau_p$ can be written as a perfect square of a sum of 2-cycle volume moduli if $k_{ppp}\neq 0$ (or trivially if $k_{pij} = 2 \sqrt{k_{pii} k_{pjj}}$ with $i\neq j$ for $k_{ppp}= 0$) since:
\be
\tau_p = \frac12 k_{pij}t^i t^j = \frac{1}{2 k_{ppp}} \left(k_{ppi}t^i\right)^2\,.
\label{square}
\ee
Out of these 232 cases, 132 are standard LVS models where $D_p$ is a dP divisor with $k_{ppp}\neq 0$, while in the remaining 100 cases $D_p$ has a different topology, and so the relation (\ref{square}) is not guaranteed to hold. In turn, in these 100 cases the volume form does not necessarily admit a simple expression in terms of 2-cycle volume moduli. Following \cite{Cicoli:2018tcq}, the 132 LVS geometries can be classified as:
\ben
\item \emph{Strong Swiss cheese}: 39 models have a volume given schematically by ${\cal V} \sim \, \tau_3^{3/2} - \, \tau_2^{3/2} - \, \tau_1^{3/2}$, which is equivalent to saying that for such models one can always find a basis where the only non-zero intersections are $k_{111}$, $k_{222}$ and $k_{333}$.

\item \emph{K3 fibrations}: 43 models are K3-fibred, leading to a volume form which can be written schematically as ${\cal V} \sim \, \tau_3\, \sqrt{\tau_2} - \, \tau_1^{3/2}$, implying that the only non-zero intersection numbers as $k_{111}$ and $k_{233}$.

\item \emph{Strong Swiss cheese-like}: 36 models feature a volume which schematically looks like $\vo \sim \, \tau_3^{3/2}  - \, (a\, \tau_1 + b\, \tau_2)^{3/2} - \, {\tau_1}^{3/2}$ where $a$ and $b$ are positive integers. This geometry is similar but qualitatively different from a strong Swiss cheese since $(a\, D_1+ b\, D_2)$ does not correspond to a smooth divisor, and so $(a\, \tau_1 + b\, \tau_2)$ cannot be redefined as a new $\tau_x$. These geometries have been used in \cite{Blumenhagen:2012kz} to study poly-instanton effects and in \cite{Cicoli:2011ct, Cicoli:2012tz, Blumenhagen:2012ue, Gao:2013hn, Gao:2014fva, Kobayashi:2017jeb} for cosmological applications.

\item \emph{Structureless}: 14 models, despite admitting a diagonal dP, do not feature a volume which can be written in terms of the $\tau$'s in a simple way. In these cases the volume can be generically expressed as $\vo \sim f_{3/2}(\tau_2,\tau_3) -\tau_1^{3/2}$ where $f$ is a homogeneous function of degree $3/2$. The relevant data for these 14 examples are presented in Tab. \ref{tab_cydata-h11eq3-14-hard-LVS} in App. \ref{AppF}. In what follows we shall focus on model $M_{3,1}$ to illustrate our numerical analysis. 
\een

\item 73 models do not admit a divisor like $D_p$ which obeys the condition (\ref{Dpcondition}), implying that in these cases the volume does not admit a simple form in terms of 4-cycle volumes. These cases are definitely of interest for our numerical study, and so in what follows we shall consider a representative example for this class of CY models, which we call $M_{3,15}$, characterised by the following topological data: 
\bea
M_{3,15}: \quad \chi &=& - \, 240\,, \quad k_{111} = 4\,, \quad k_{112} = -2\,, \quad k_{113} = -2\,, \quad k_{123} = 2\,, \nn \\
&& \text{K\"ahler cone:} \quad t_1 > 0\,, \quad  t_2 > t_1\,, \quad t_3 > t_1\,.
\label{M315data}
\eea
\ei 
Let us finally point out that this discussion shows that 118 models out of 132 LVS models can definitely be studied using the conventional approach based on 4-cycle moduli, corresponding to 38.7\% of all 305 cases. However the 14 `structureless' LVS models and the 73 models without a divisor which satisfies (\ref{Dpcondition}) are certainly better analysed using 2-cycle moduli, corresponding to 28.5\% of all 305 cases.

\section{Moduli stabilisation results}
\label{NewIIBVacua}

\subsection{Reproducing old vacua}
\label{KKLTstab}

As a warmup to check the validity of our numerical analysis, we first focus on standard KKLT vacua with a single K\"ahler modulus and typical LVS models with 2 K\"ahler moduli one of which is a diagonal dP divisor.

\subsubsection{KKLT with a single modulus}

The potential of the simplest KKLT model with a single K\"ahler modulus and no $\alpha'$-corrections is given by our master formula (\ref{MasterF}) which for $h^{1,1} = n=1$ and $\hat\xi=0$ reduces to (setting $\la s \ra = g_s^{-1}$):
\be
V_{\rm KKLT} = \frac{9 k_{111} g_s\, e^{K_{\rm cs}}}{\tau_1^2}\,a_1 |A_1|\, e^{-a_1 \tau_1} \left[ |W_0| \, \cos\left(a_1\, \rho_1 + \theta_0 - \phi_1\right) + \frac{|A_1|}{3} \, e^{-a_1 \tau_1} \left(a_1\, \tau_1+3 \right)\right].
\label{VKKLT}
\ee
After performing the axion minimisation by setting $(a_1 \rho_1 + \theta_0 - \phi_1) = \pi$, a simple calculation leads to the following relation in a generic extremum for the saxion $\tau_1$:
\be
V_0 \equiv \la V_{\rm KKLT} \ra =  -\frac{3 k_{111}g_s\, e^{K_{\rm cs}}}{\la\tau_1\ra} \, a_1^2 |A_1|^2 e^{-2 a_1 \la\tau_1\ra}\leq 0\,,
\label{NoGoKKLT}
\ee
which excludes dS vacua in the minimal KKLT model. This problem can be circumvented by adding uplifting contributions which can come from several different sources. In the case of anti-branes, the uplifting term can be simply written as:
\be
V_{\rm KKLT}^{\rm up} =  V_{\rm KKLT} + V_{\rm up}\qquad \text{with}\qquad V_{\rm up} = \frac{\delta}{\tau_1^p}\,,
\label{eq:kklt-up}
\ee
where $\delta > 0$ is a tunable flux dependent parameter. The new term $V_{\rm up}$ modifies the condition in (\ref{NoGoKKLT}) as follows:
\be
V_0^{\rm up} \equiv \la V_{\rm KKLT}^{\rm up}\rangle = V_0 +  \la V_{\rm up}\ra \left(1 - \frac{p}{a_1\la \tau_1\ra + 2} \right),
\ee
showing that the dS no-go condition can be avoided for $0 < p < a_1\la \tau_1\ra + 2$. Using Tab. \ref{tab_cydata-h11eq1}, the potentials (\ref{VKKLT}) and (\ref{eq:kklt-up}), where we have set $p=3$, can be minimised numerically for all 5 CY threefolds with $h^{1,1}=1$, leading to the results given in Tab. \ref{tab_min-h11eq1}. One can easily check that all minima lie in a region where the effective field theory is under control since each condition of Sec. \ref{EFT} is satisfied.

\begin{center}
\begin{tabular}{|c||c|c|c||c|c|c|c||} 
\hline
Model & $\la t_1\ra$ & $\la\vo\ra$ & $V_0 \times 10^{15}$ & $\delta \times 10^8$ & $\la t_1\ra$& $\la\vo\ra$ &$V_0^{\rm up} \times 10^{15}$ \\
\hline
$M_{1,1}$ & 15.0724 & 570.688 & $-3.97181$ & $5$ & 15.1637 & 581.12 & 0.220944 \\
$M_{1,2}$ & 6.7406 & 255.220 & $-19.8590$ & $0.3$ & 6.80975 & 263.155 & 11.265 \\
$M_{1,3}$ & 8.70207 & 329.487 & $-11.9154$ & $1$ & 8.81984 & 343.046 & 10.3356\\
$M_{1,4}$ & 15.0724 & 570.688 & $-3.97181$ & $5$ & 15.1637 & 581.12 & 0.220944 \\
$M_{1,5}$ & 10.6578 & 403.537 & $-7.94361$ & $2$ & 10.7779 & 417.331 & 5.30588 \\
\hline
\end{tabular}
\captionof{table}{KKLT vacua with and without anti-brane uplifting for all CY threefolds with $h^{1,1}=1$. The underlying parameters are set as $W_0 = - 10^{-4}$, $a_1 = 0.1$, $g_s = 0.1$, $K_{\rm cs} = 0.1$, $A_1 = 1$ and the axion $\rho_1$ is minimised at $\la\rho_1\ra = 0$.}
\label{tab_min-h11eq1}
\end{center}

\begin{figure}[H]
\begin{center}
\includegraphics[width=10.0cm,height=7.0cm]{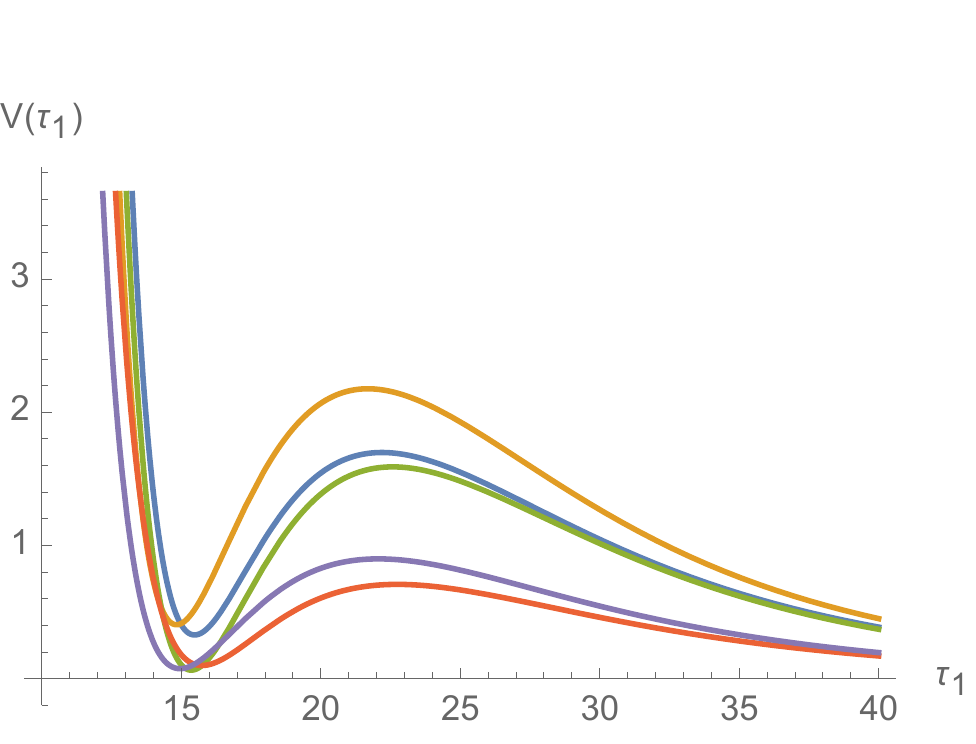}
\end{center}
\caption{KKLT scalar potentials $V(\tau_1)$ ($\times 10^5$) with $\alpha'$-uplift for all 5 models in Tab. \ref{tab_alphamin-h11eq1}.}
\label{fig_upKKLT}
\end{figure}

In the KKLT framework, dS vacua can also be achieved by including $\alpha'$ corrections which are captured by our master formula (\ref{MasterF}) for $\hat\xi\neq 0$. In this case, the scalar potential is the sum of the following 3 terms (after fixing again $(a_1 \rho_1 + \theta_0 - \phi_1) = \pi$):
\bea
V_{\mathcal{O}(\alpha'^3)} &=&  e^{{\cal K}} \, \frac{3 \, \hat\xi (\vo^2 + 7\, \hat\xi \,\vo +\hat{\xi}^2)}{(\vo-\hat\xi) (2\vo+ \hat\xi)^2}\, \,|W_0|^2,\\
V_{\rm np1} &=& - 2\, e^{{\cal K}}\,|W_0| \, |A_1|\, e^{- a_1 \tau_1} \left[\frac{(4\vo^2  + \vo \, \hat\xi + 4\, \hat{\xi}^2)}{(\vo - \hat\xi) (2 \vo + \hat\xi)}\, (a_1 \tau_1) +\frac{3 \, \hat\xi (\vo^2+ 7\, \hat\xi \,\vo +\hat{\xi}^2 )}{(\vo-\hat\xi) (2\vo + \hat\xi )^2}\right ],\nn\\
V_{\rm np2} &=& e^{{\cal K}}\,  \, |A_1|^2 \, e^{- 2 a_1 \tau_1} \biggl[ -4\, a_1^2 \left(\vo+\frac{\hat{\xi}}{2}\right) \, \sqrt{2\, k_{111}\,\tau_1}  + \frac{4\vo - \hat\xi}{(\vo - \hat\xi)} (a_1 \tau_1)^2 \nn \\
&+& \frac{(4 \vo^2 + \vo \, \hat{\xi} + 4\, \hat{\xi}^2)}{ (\vo - \hat\xi) (2 \vo + \hat\xi)}\, (2\,a_1\,\tau_1) +\frac{3 \, \hat\xi (\vo^2  + 7\, \hat\xi \,\vo + \hat{\xi}^2)}{(\vo-\hat\xi) (2 \vo + \hat\xi)^2}\biggr]. \nn
\eea
This scalar potential can be minimised numerically with respect to either $t_1$ or $\tau_1$, since in this simple case the conversion between 2- and 4-cycle moduli is trivial. The results of our numerical analysis are presented in Tab. \ref{tab_alphamin-h11eq1} which shows that these dS vacua tend to be located at relatively small volume. However the effective field theory is still marginally under control since all conditions listed in Sec. \ref{EFT} are satisfied. In particular, the condition (\ref{cond4}) is still slightly met even for relatively small values of $\vo$ since $m_{3/2}\simeq 0.1 \,M^{({\rm bulk})}_{\rm KK}$. For a condensing gauge group with rank $N=32$, (corresponding to $a_1 = \pi/16$), $\tau_1$ at the dS minimum is around $15$ as can be seen from Fig. \ref{fig_upKKLT}. Larger values of $\la \vo\ra$ of $\mathcal{O}(50 -100)$ can be realised for $N\sim\mathcal{O}(100)$ and $|W_0| \sim \mathcal{O}(50-100)$ \cite{Balasubramanian:2004uy, Westphal:2006tn, Rummel:2011cd, Louis:2012nb} even if these values have their own limitations and criticism \cite{Cicoli:2013swa, Conlon:2012tz}.

\begin{center}
\begin{tabular}{|c||c|c|c|c|c|c|c|} 
\hline
Model & $g_s$ & $-W_0$ & $\la t_1\ra$ & $\la\tau_1\ra$ & $\la\vo\ra$ & $\hat\xi$ & $V_0 \times 10^7$ \\
\hline
$M_{1,1}$ & 0.1 & 4.55 & 5.56456 & 15.4822  & 28.7172 & 1.53245 &  0.327441  \\
$M_{1,2}$ & 0.2 & 0.68 & 2.43598 & 14.8350 & 12.0459 & 2.70901 & 0.404328 \\
$M_{1,3}$ & 0.2 & 0.93 & 3.20034 & 15.3633 & 16.3892 & 2.76319 &  6.30582 \\
$M_{1,4}$ & 0.2 &1.24 & 5.61173 & 15.7457 & 29.4536 & 3.90097 & 9.80739 \\
$M_{1,5}$ & 0.2 & 0.74 & 3.86787 & 14.9604 & 19.2883 & 4.00933 & 7.42126 \\
\hline
\end{tabular}
\captionof{table}{dS KKLT vacua with $\alpha'$-uplift for all CYs with $h^{1,1}=1$. The underlying parameters are set as $a_1 = \frac{\pi}{16}$, $K_{\rm cs} = 1$, $A_1 = 1$ and the axion $\rho_1$ is minimised at $\la\rho_1\ra = 0$.}
\label{tab_alphamin-h11eq1}
\end{center}

\subsubsection{LVS with diagonal del Pezzo}

The potential of the simplest LVS model with 2 K\"ahler moduli and volume of the form $\vo\simeq \tau_2^{3/2}-\tau_1^{3/2}$, can be obtained from our master formula (\ref{MasterF}) setting $h^{1,1} = 2$, $n=1$ and $\hat\xi\neq 0$. Taking the large volume limit and minimising the axionic direction $\rho_1$, this scalar potential can be approximated as the sum of 3 terms:
\be
V_{\rm LVS} = V_1 + V_2 + V_3\,,
\label{eq:lvs3term-pot}
\ee
with:
\be
V_1 = \frac{\alpha\, \sqrt{\tau_1}\, e^{-2\, a_1\, \tau_1} }{\vo}>0\,, \qquad V_2 = - \frac{\beta\, \tau_1 \, e^{- a_1\, \tau_1}}{\vo^2}<0 \,, \qquad V_3 = \frac{\gamma}{\vo^3}\,,
\ee
where the model-dependent parameters $\alpha$ and $\beta$ are positive while the sign of $\gamma\propto -\chi$ depends on the sign of the CY Euler number $\chi$. Notice that the minus sign in front of $V_2$ is due to the $\rho_1$ minimisation for $\beta >0$. The 3 terms in (\ref{eq:lvs3term-pot}) are of the same order if $\vo\sim e^{a_1\tau_1}$, and so any extremum of this potential lies at exponentially large volume.

Let us now analyse the vacuum structure of the LVS potential (\ref{eq:lvs3term-pot}). Trading $\tau_2$ for the overall volume $\vo$, the extremisation conditions $\partial_\vo V_{\rm LVS}  = \partial_{\tau_1} V_{\rm LVS} = 0$ lead to the following relations among the 3 terms of the LVS potential at any extremum:
\be
\la V_1\ra = \la V_3\ra \left(1 - \frac{1}{a_1 \la\tau_1\ra} \right)>0\,, \qquad \la V_2\ra = -\la V_3\ra \left(2 - \frac{1}{2\, a_1 \la\tau_1\ra} \right)<0\,,
\label{eq:lvs-extremum}
\ee
which imply that at any extremum:
\be
\la V_{\rm LVS}\ra = - \frac{\la V_3\ra}{2\, a_1 \la\tau_1\ra}\,.
\label{Vextremum}
\ee
In the regime where the instanton series is under control, i.e. for $a_1 \la\tau_1\ra \gg 1$, the second expression in (\ref{eq:lvs-extremum}) simplifies to $\la V_2\ra = -\la V_3\ra <0$ which implies that a solution can exist only if $\la V_3\ra > 0$, i.e. for negative CY Euler number since $\gamma\propto -\chi$. In turn, (\ref{Vextremum}) forces the potential to be negative at any extremum, i.e. $\la V_{\rm LVS}\ra < 0$. This is a no-go result for any dS extremum, both potential minima and maxima. This is consistent with the known fact that LVS models without any uplifting term give rise just to AdS minima.

This no-go result can be evaded by adding an additional positive contribution to the LVS potential (\ref{eq:lvs3term-pot}) which can be expressed as $V_4 = \delta/\vo^p$ with $\delta >0$. This term can come from either anti-D3s \cite{Kachru:2003aw}, non-perturbative effects at singularities \cite{Cicoli:2012fh} or T-branes \cite{Cicoli:2015ylx}, and modifies the relation (\ref{Vextremum}) for the value of the potential $V = V_{\rm LVS} + V_4$ at any extremum as follows: 
\be
\la V\ra = - \, \frac{\la V_3\ra}{2\, a_1 \la\tau_1\ra} + \la V_4\ra \left(1 - \frac{p}{3} - \frac{p}{6 a_1\la\tau_1\ra} \right).
\label{VLVSup}
\ee
For $p = 3$, $V_4$ can be reabsorbed into $V_3$ via a proper shift of $\gamma$ to include $\delta$, and so $\la V\ra$ is still negative. More generally, one can see that $\la V\ra<0$ for $p \geq 3$. Hence, the only way to evade the dS no-go result found above is to consider $p < 3$. In this case, (\ref{VLVSup}) shows clearly that one can easily obtain $\la V \ra >0$ since the term proportional to $\la V_4\ra$ becomes positive (for $a_1\la\tau_1\ra \gg 1$) and can compensate the fact that the term proportional to $\la V_3\ra$ is negative. This is natural since in order to obtain dS minima the uplifting term has to dominate at large volume.

\subsection{Discovering new vacua}
\label{NewVacua}

Let us now show how new classes of type IIB vacua can be found with the help of numerical techniques and minimising the scalar potential with respect to 2-cycle volume moduli.

\subsubsection{New KKLT vacua}

Let us now focus on a case with $h^{1,1}=2$, which leads to new KKLT vacua. This is model $M_{2,6}$ in Tab. \ref{tab_cydata-h11eq2} in App. \ref{AppE} which we classify as `hard' since $\vo$ does not admit a simple form in terms of the 4-cycle volume moduli. In this case we shall perform numerical moduli stabilisation using the 2-cycle volume moduli.

The CY threefold of model $M_{2,6}$ corresponds to the polytope ID $\#10$ in the CY database of \cite{Altman:2014bfa} and it is defined by the following toric data:
\begin{center}
\begin{tabular}{|c|cccccc|}
\hline
    CY & $x_1$  & $x_2$  & $x_3$  & $x_4$  & $x_5$ & $x_6$      \\
    \hline
3 & 1 & 1  & 1 & 0 & 0 & 0  \\
3 & 0 & 0  & 0 & 1 & 1 & 1  \\
\hline
 & SD$_1$  & SD$_1$ & SD$_1$ &  SD$_2$ & SD$_2$ & SD$_2$  \\
\hline
\end{tabular}
\end{center}
where what we call the `special deformation' divisors SD$_1$ and SD$_2$ (following the nomenclature of \cite{Gao:2013pra, Cicoli:2016xae, Cicoli:2017axo}) are represented by the following Hodge diamond:
\bea
& & {\rm SD}_1 \equiv \begin{tabular}{ccccc}
    & & 1 & & \\
   & 0 & & 0 & \\
  2 & & 30 & & 2 \\
   & 0 & & 0 & \\
    & & 1 & & \\
  \end{tabular} \equiv {\rm SD}_2\,.
\eea
The Hodge numbers are $(h^{2,1}, h^{1,1}) = (83, 2)$, the Euler number is $\chi=-162$ and the SR ideal is ${\rm SR} =  \{x_1 \, x_2 \, x_3, \,  \, x_4 \, x_5 \, x_6\}$. As can be seen from Tab. \ref{tab_cydata-h11eq2} in App. \ref{AppE}, the intersection numbers and the K\"ahler cone in the basis of smooth divisors $D_1=\{1,0\}={\rm SD}_1$ and $D_2 =\{0,1\} = {\rm SD}_2$ are:
\be
k_{111} = k_{222} = 0\,, \qquad  k_{112} = k_{122} = 3\,,   \qquad \text{K\"ahler cone:} \quad t_1 > 0\,, \, \, t_2 > 0\,. 
\ee
In this case the overall volume and the 4-cycle moduli take the following form:
\be
\vo = \frac32\left(t_1^2\, t_2+ t_1\, t_2^2\right), \qquad \tau_1 = 3\, t_1 t_2 + \frac32\,t_2^2\,, \qquad \tau_2 = 3\, t_1 t_2 + \frac32\,t_1^2\,. \nn
\ee
The scalar potential of this non-standard KKLT model can be obtained from our master formula (\ref{MasterF}) by setting $h^{1,1} = n = 2$ and $\hat\xi=12.4128$ which follows from $\chi=-162$ and our choice of the string coupling $g_s=0.1$. 

Notice that, under the exchange $t_1 \leftrightarrow t_2$, both $\vo$ and the K\"ahler cone are invariant and $\tau_1 \leftrightarrow \tau_2$. Thus the scalar potential is symmetric if $A_1=A_2$ and $a_1=a_2$. In this case, the minimisation solutions have therefore to be symmetric. In Tab. \ref{tab_min-Model-A} we present such symmetric solutions, along with non-symmetric ones for $A_1 \neq A_2$ and/or $a_1 \neq a_2$. We only show vacua which are AdS, but larger values of $|W_0|$ would give rise to $\alpha'$-uplifted dS solutions. All minima lie in a region where the effective 4D theory is under control.

\begin{center}
\begin{tabular}{|c|c|c|c|c|c|c|c|c||} 
\hline
$W_0$ & $\{A_1, A_2\}$ & $a_1$ & $a_2$ & $\la t_1\ra$ & $\la t_2\ra$ & $\la\vo\ra$ & $V_0\,(\times 10^9)$ \\
\hline
$-$0.01 & \{100, 100\} & $\pi/8$ & $\pi/8$ & 2.62890 & 2.62890  & 54.5059 & $-$8.97384 \\
$-$0.01 & \{50, 50\} & $\pi/10$ & $\pi/10$  & 2.84489 & 2.84489 & 69.0745 & $-$5.73264 \\
$-$0.10  & \{50, 50\} & $\pi/16$ & $\pi/16$ & 3.17337 & 3.17337 & 95.8698 & $-$291.656 \\
$-$0.01 & \{100, 100\} & $\pi/12$ & $\pi/8$ & 2.07977 & 3.89165 & 72.4963 & $-$5.30985 \\
$-$0.01 & \{170, 180\} & $\pi/12$ & $\pi/11$ & 3.01803 & 3.43224 & 100.224 & $-$2.91829 \\
\hline
\end{tabular}
\captionof{table}{Benchmark examples for Model $M_{2,6}$.}
\label{tab_min-Model-A}
\end{center}

\begin{figure}[H]
\begin{center}
\includegraphics[width=14.0cm,height=7.0cm]{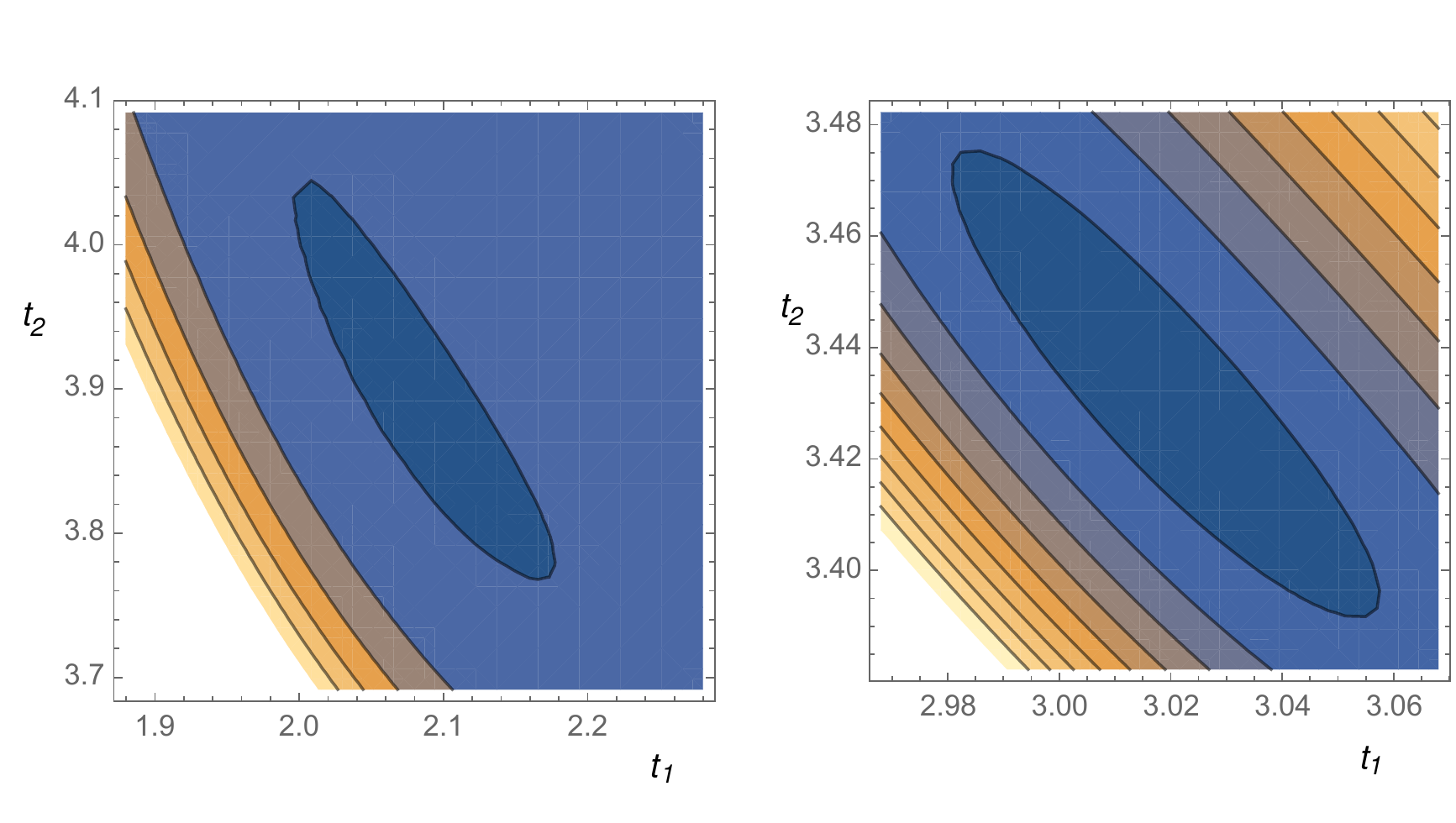}
\end{center}
\caption{Contour plots of the potential in the $(t_1, t_2)$-plane for the last two models of Tab. \ref{tab_min-Model-A}.}
\label{fig_A}
\end{figure}

Larger values of the CY volume can be realised by increasing the rank of the condensing gauge group, which may be beneficial to gain better control over the effective theory. In Fig. \ref{fig_A} we present contour plots for the scalar potential in the $(t_1, t_2)$-plane, showing the AdS minimum of the last two cases presented in Tab. \ref{tab_min-Model-A}.

\subsubsection{New LVS vacua with a diagonal and a non-diagonal del Pezzo}

We now focus on a new LVS model with $h^{1,1}=3$ which we classified as `structureless' since $\vo$ does not admit a simple form in terms of 4-cycle moduli. This is model $M_{3,1}$ in Tab. \ref{tab_cydata-h11eq3-14-hard-LVS} in App. \ref{AppF}. In this case we will stabilise the moduli numerically using the 2-cycle moduli.

\begin{figure}[H]
\begin{center}
\hspace*{-0.5cm} \includegraphics[width=15.0cm,height=4.0cm]{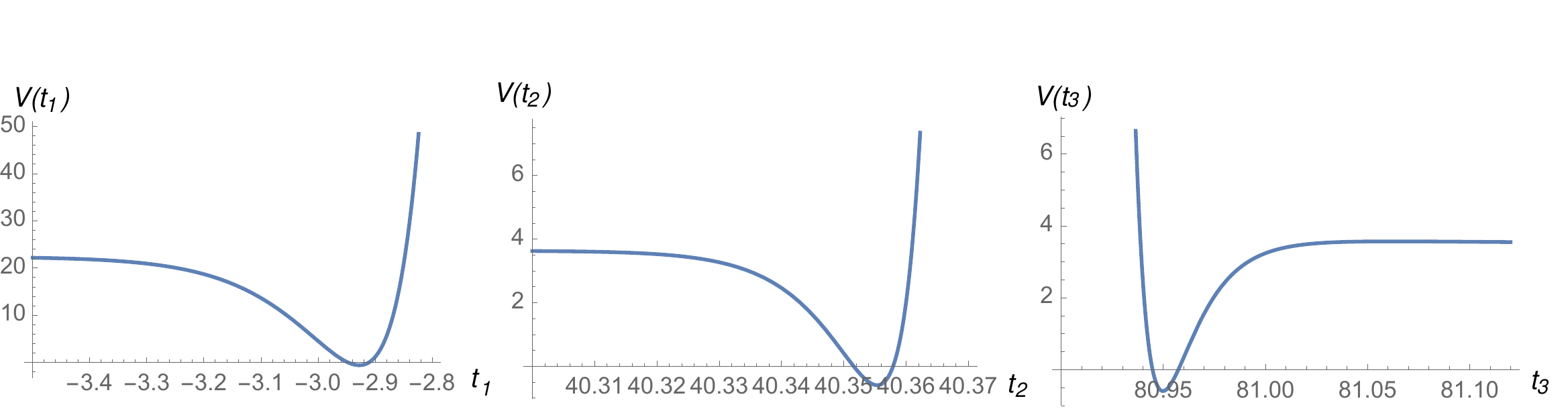}
\end{center}
\caption{Scalar potential for each of the 3 K\"ahler moduli $(t_1, t_2, t_3)$ (with the other 2 fixed at their minima) for example E6 of Tab. \ref{tab_min-Model-C}.}
\label{fig_B}
\end{figure}

The CY threefold of model $M_{3,1}$ corresponds to the polytope ID $\#61$ in the CY database of \cite{Altman:2014bfa} and it is defined by the following toric data:
\begin{center}
\begin{tabular}{|c|ccccccc|}
\hline
& $x_1$  & $x_2$  & $x_3$  & $x_4$  & $x_5$ & $x_6$  & $x_7$       \\
\hline
7 & 0  & 2 & 1 & 1 & 0 & 1  & 2   \\
4 & 1  & 0 & 0 & 0 & 1 & 1  & 1   \\
3 & 1  & 1 & 0 & 0 & 0 & 0  & 1   \\   
\hline
& dP$_1$  & NdP$_{22}$ & SD$_1$ &  SD$_1$ & dP$_8$ & SD$_2$  &  SD$_3$  \\
\hline
\end{tabular}
\end{center}
The Hodge numbers are $(h^{2,1}, h^{1,1}) = (66, 3)$, the Euler number is $\chi=-126$ and the SR ideal is:
\be
{\rm SR} =  \{x_1 x_5, \, x_5 x_6, \, x_1 x_2 x_7, \, x_3 x_4 x_6, \, x_2 x_3 x_4 x_7 \} \,. \nn
\ee
The analysis of the divisor topologies shows that they can be represented by the following Hodge diamonds:
\bea
{\rm dP}_n \, \, {\rm or} \, \,  {\rm NdP}_n &\equiv& \begin{tabular}{ccccc}
    & & 1 & & \\
   & 0 & & 0 & \\
  0 & & n+1 & & 0 \\
   & 0 & & 0 & \\
    & & 1 & & \\
  \end{tabular}, \qquad \quad {\rm SD}_1 \equiv \begin{tabular}{ccccc}
    & & 1 & & \\
   & 0 & & 0 & \\
  1 & & 21 & & 1 \\
   & 0 & & 0 & \\
    & & 1 & & \\
  \end{tabular}, \nn \\
{\rm SD}_2 &\equiv& \begin{tabular}{ccccc}
    & & 1 & & \\
   & 0 & & 0 & \\
  2 & & 30 & & 2 \\
   & 0 & & 0 & \\
    & & 1 & & \\
  \end{tabular}, \qquad \qquad {\rm SD}_3 \equiv \begin{tabular}{ccccc}
    & & 1 & & \\
   & 0 & & 0 & \\
  4 & & 44 & & 4 \\
   & 0 & & 0 & \\
    & & 1 & & \\
  \end{tabular}. \nn
\eea
As can be seen from Tab. \ref{tab_cydata-h11eq3-14-hard-LVS}, the intersection numbers and the K\"ahler cone in the basis of smooth divisors $D_1=\{0,1,0\}={\rm dP}_8$, $D_2 = \{0,1,1\}={\rm dP}_1$ and $D_3=\{1,1,0\}={\rm SD}_2$ are:
\bea
k_{111} &=& 1, \quad k_{222} = 8\,, \quad k_{223} = -5\,, \quad k_{233} = 3\,, \quad k_{333} = 0\,,  \nn \\
\text{K\"ahler cone:} &&  t_1< 0\,, \quad t_3 -2 t_2 > 0\,, \quad t_1 + t_3 > 0\,, \quad t_1 + 3 \, t_2 > 0\,,
\label{KahCone}
\eea
which shows clearly that $D_1$ is a diagonal dP$_8$ while $D_2$ is a non-diagonal dP$_1$. In addition, the overall volume and the 4-cycle volume moduli are as follows:
\bea
\vo &=& \frac16\,t_1^3 + \frac16\left(8\, t_2^3 - 15\, t_2^2\, t_3 + 9\, t_2\, t_3^2\right), \nn \\
\tau_1 &=& \frac12\,t_1^2\,, \qquad \tau_2 = 4 \, t_2^2 - 5\, t_2\, t_3 + \frac32\, t_3^2\,, \qquad \tau_3 = 3\,t_2\, t_3- \frac52\, t_2^2 \,.
\label{Taus}
\eea
This shows clearly that $\vo$ does not admit a simple expression in terms of 4-cycle moduli. The potential of this structureless LVS model can be obtained from our master formula (\ref{MasterF}) by setting $h^{1,1} = 3$, $n = 2$ and $\hat\xi\neq 0$. In Tab. \ref{tab_min-Model-C} we present the results of our numerical minimisation with respect to 2-cycle moduli for different choices of the microscopic parameters, while Fig. \ref{fig_B} shows the minima for each of the 3 $t$-moduli for a particular example (E6 in Tab. \ref{tab_min-Model-C}). All minima are within the regime of validity of the effective theory.
 
\begin{center}
\begin{tabular}{|c||c|c|c|c|c|c|c|c|c|c||} 
\hline
Example & $g_s$ & $\la t_1\ra$ & $\la t_2\ra$ & $\la t_3\ra$ & $\la\vo\ra$ & $\hat\xi$ & $V_0$ \\
\hline
E1 & 0.15 & $-$2.22639 & 5.06509 & 10.9593 & 381.042 & 5.2552 & $- 1.13561\times 10^{-9}$ \\
E2 & 0.14 & $-$2.29369 & 6.12009 & 13.0085 & 639.008 & 5.8282 & $-1.93895 \times 10^{-10}$ \\
E3 & 0.13 & $-$2.38288 & 7.87495 & 16.4338 & 1291.22 & 6.51345 & $-1.79005 \times 10^{-11}$ \\
E4 & 0.12 & $-$2.51017 & 11.3404 & 23.243 & 3658.79 & 7.34436 & $-5.73649 \times 10^{-13}$ \\
E5 & 0.11 & $-$2.69616 & 19.6372 & 39.673 & 18208.5 & 8.36829 & $-3.67679 \times 10^{-15}$ \\
E6 & 0.10 & $-$2.92776 & 40.3554 & 80.9498 & 154711.0 & 9.65442 & $-5.94805 \times 10^{-18}$ \\
\hline
\end{tabular}
\captionof{table}{Benchmark examples for model $M_{3,1}$ where we have set $K_{\rm cs} = A_1 = A_2 = 1$, $W_0 = -1$, $a_1 = \pi$, $a_2 = \pi/2$ and $A_3 = 0$.}
\label{tab_min-Model-C}
\end{center}

As can be seen from Tab. \ref{tab_min-Model-C}, smaller values of the string coupling give rise to larger values of $\vo$ which improve the control over the effective field theory. This is expected since it resembles the behaviour of standard LVS where the small 2-cycle $t_1$ and the volume $\vo$ would scale respectively as $\la t_1\ra \propto g_s^{-1/2}$ and $\la\vo\ra \propto e^{\la t_1\ra^2}$. 

A qualitative understanding of the results of our numerical minimisation can be gained as follows. First note that, given that dP$_8$ is a diagonal divisor, $\mathcal{O}(\alpha'^3)$ corrections to $K$ and a single non-perturbative effect in $W$ with $A_1 \neq 0$, would be sufficient to fix the volume exponentially large and $\la t_1\ra\sim g_s^{-1/2}$ together with the axion $\rho_1$. This would however leave 3 flat directions which can be parametrised for example by $t_2$, $\rho_2$ and $\rho_3$. Because of the axionic shift symmetry, the 2 flat directions $\rho_2$ and $\rho_3$ can be lifted only by $T_2$- and $T_3$-dependent non-perturbative corrections to $W$ which would generate a potential also for $t_2$. However $t_2$ can develop a potential also via perturbative corrections to $K$ which would in general dominate over non-perturbative effects if $\tau_2$ and $\tau_3$ (the 2 combinations appearing in the exponent of the non-perturbative superpotential) were both large cycles. 

Examples where the remaining saxionic flat direction is fixed by higher order $\alpha'$ corrections are given in \cite{Cicoli:2016chb}. However our master formula (\ref{MasterF}) does not include this kind of effect, and so we looked for minima where the remaining saxionic flat direction is fixed at small values, so that the dominant source of its potential is non-perturbative physics. This is possible for $A_2 \neq 0$ and $A_3=0$ if the moduli are stabilised close to the K\"ahler cone condition $t_3>2 t_2$ in (\ref{KahCone}), but still far enough from it to be able to trust the effective field theory. In fact, for $t_3 \to 2 t_2$, the expressions (\ref{Taus}) show that $\tau_2 \to 0$ and $\tau_3 \to \tfrac72 t_2^2$, and so a superpotential contribution of the form $A_2\,e^{-a_2 T_2}$ would not be suppressed, while $A_3\,e^{-a_3 T_3}$ would be negligible for $t_2$ stabilisation even if it would be crucial to fix $\rho_3$.
Tab. \ref{tab_min-Model-C1} shows that the minima displayed in Tab. \ref{tab_min-Model-C} are still fully within the regime of validity of the effective field theory since each 4-cycle modulus, in particular $\tau_2$, turns out to be fixed at values much larger than unity. 

\begin{center}
\begin{tabular}{|c||c|c|c|c|} 
\hline
Example  & $\la\tau_1\ra$ & $\la\tau_2\ra$ & $\la\tau_3\ra$ & $\la\vo\ra$ \\
\hline
E1 & 2.47842 & 5.23087 & 102.392 & 381.042 \\
E2 & 2.63050 & 5.58806 & 145.201 & 639.008 \\
E3 & 2.83905 & 6.08695 & 233.208 & 1291.22 \\
E4 & 3.15047 & 6.84907 & 469.243 & 3658.79 \\
E5 & 3.63465 & 8.06464 & 1373.15 & 18208.5 \\
E6 & 4.28590 & 9.73469 & 5728.89 & 154711.0 \\
\hline
\end{tabular}
\captionof{table}{Values of the 4-cycle moduli for the benchmark examples of model $M_{3,1}$ listed in Tab. \ref{tab_min-Model-C}.}
\label{tab_min-Model-C1}
\end{center}

In the general case where also $A_3\neq 0$ the scalar potential should feature two minima: ($i$) an LVS-like AdS vacuum with the same properties described just above but with a massive, even if ultra-light, $\rho_3$ axion; and ($ii$) an $\alpha'$-uplifted KKLT-like dS minimum where however the CY volume would take values smaller than the one shown in Tab. \ref{tab_min-Model-C} which would affect the trustability of the effective field theory.

Let us finally point out that our numerical analysis has shown how model $M_{3,1}$ can lead to an LVS-like vacuum where not only the diagonal dP$_8$ modulus $\tau_1$, but also the non-diagonal dP$_1$ modulus $\tau_2$, can be fixed at `small' size. This observation raises the question of whether it is possible to find new LVS vacua for CY threefolds which do not admit diagonal divisors. This issue is addressed in the next section.

\subsubsection{New LVS vacua without a diagonal del Pezzo}
\label{NewLVS2}

\begin{figure}[H]
\begin{center}
\hspace*{-0.5cm} \includegraphics[width=15.0cm,height=4.0cm]{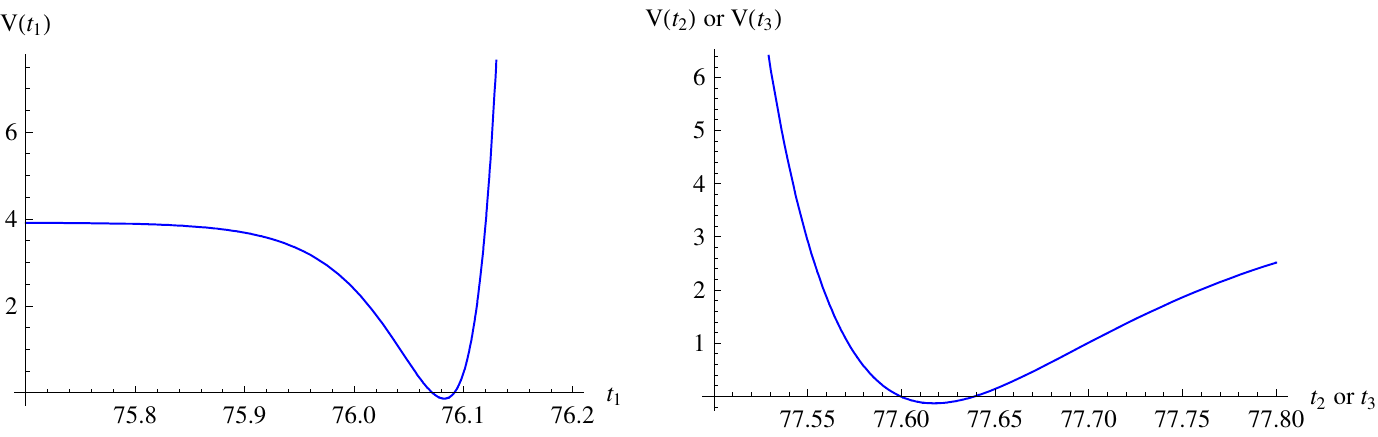}
\end{center}
\caption{Scalar potential $V$ ($\times 10^{17}$) for each of the K\"ahler moduli $(t_1, t_2, t_3)$ (with the other two fixed at their minima) for example E6 of Tab. \ref{tab_min-Model-D}.}
\label{fig_Bo}
\end{figure}

In this section we discuss a 3-moduli CY model that does not feature any diagonal dP divisor. We shall  show that despite this an LVS-like AdS minimum at exponentially large volume still exists thanks to a particular symmetry of the K\"ahler moduli space. We shall perform a detailed analysis of K\"ahler moduli stabilisation both via numerical techniques and analytical approximations in terms of 2-cycle volume moduli focusing on model $M_{3,15}$ introduced in Sec. \ref{h113}. The CY threefold of model $M_{3,15}$ corresponds to the polytope ID $\#263$ in the CY database of \cite{Altman:2014bfa} and it is defined by the following toric data:
\begin{center}
\begin{tabular}{|c|ccccccc|}
\hline
& $x_1$  & $x_2$  & $x_3$  & $x_4$  & $x_5$ & $x_6$  & $x_7$       \\
\hline
6 & 0      & 0  & 1  & 1  & 0  & 1      & 3   \\
6 & 0      & 1  & 0  & 0  & 1  & 1      & 3   \\
4 & 1      & 0  & 0  & 0  & 0  & 1      & 2   \\   \hline
  & dP$_5$ & K3 & K3 & K3 & K3 & SD$_1$ & SD$_2$  \\
\hline
\end{tabular}
\end{center}
The Hodge numbers are $(h^{2,1}, h^{1,1}) = (123, 3)$, the Euler number is $\chi=-240$ and the SR ideal is ${\rm SR} =  \{x_1 x_6, \, x_2 x_5,\, x_3 x_4 x_7 \}$. The analysis of the divisor topologies shows that they are represented by the following Hodge diamonds:
\bea
& & {\rm dP}_5 \equiv \begin{tabular}{ccccc}
    & & 1 & & \\
   & 0 & & 0 & \\
  0 & & 6 & & 0 \\
   & 0 & & 0 & \\
    & & 1 & & \\
  \end{tabular}, \qquad \qquad {\rm K3} \equiv \begin{tabular}{ccccc}
    & & 1 & & \\
   & 0 & & 0 & \\
  1 & & 20 & & 1 \\
   & 0 & & 0 & \\
    & & 1 & & \\
  \end{tabular}, \nn \\
& & {\rm SD}_1 \equiv \begin{tabular}{ccccc}
    & & 1 & & \\
   & 0 & & 0 & \\
  4 & & 46 & & 4 \\
   & 0 & & 0 & \\
    & & 1 & & \\
  \end{tabular}, \qquad \quad {\rm SD}_2 \equiv \begin{tabular}{ccccc}
    & & 1 & & \\
   & 0 & & 0 & \\
  29 & & 196 & & 29 \\
   & 0 & & 0 & \\
    & & 1 & & \\
  \end{tabular}. \nn
\eea
As can be seen from (\ref{M315data}), the intersection polynomial in the basis of smooth divisors $D_1 = \{0, 0, 1\}={\rm dP}_5$, $D_2 = \{0, 1, 0\} = {\rm K3}$ and $D_3 = \{1, 0, 0\} = {\rm K3}$ is given by:
\be
I_3 = 4\, D_1^3 - 2 \,D_1^2\, D_2 - 2\, D_1^2\, D_3 + 2\, D_1\,D_2\,D_3\,.
\label{I3a}
\ee
The linearity of $I_3$ in $D_2$ and $D_3$, together with the divisor analysis, shows that this CY threefold is K3-fibred. Moreover the fact that $k_{112}$, $k_{113}$ and $k_{123}$ are all non-zero implies that $D_1$ is a non-diagonal dP$_5$ divisor. The CY volume and the 4-cycle moduli become:
\bea
\vo &=& \frac23\, t_1^3 - t_1^2\left(t_2 + t_3\right) + 2\,t_1\,t_2\, t_3 \,, \nn \\
\tau_1 &=& 2 \left(t_1 - t_2\right)\left(t_1 - t_3\right), \qquad \tau_2 = t_1 \left(2\,t_3 - t_1\right), \qquad \tau_3 = t_1\left(2\,t_2 - t_1\right). 
\label{Topol}
\eea
The potential of this model can be obtained from (\ref{MasterF}) by setting $h^{1,1}=3$, $n=1$ and $\hat\xi>0$. In Tab. \ref{tab_min-Model-D} we present the results of our numerical minimisation with respect to 2-cycle moduli for different choices of the microscopic parameters, while Fig. \ref{fig_Bo} shows the minima for each of the 3 $t$-moduli for example E6 in Tab. \ref{tab_min-Model-D}. The large values of $\la\vo\ra$ show that this model features an LVS-like AdS vacuum even though it does not have a diagonal dP divisor. All minima lie in a region where the effective field theory is fully under control.

\begin{center}
\begin{tabular}{|c||c|c|c|c|c|c|c|c|c|c||} 
\hline
Example & $g_s$ & $N$ & $ \la t_1\ra$ & $\la t_2 \ra$ = $\la t_3\ra$ & $\la\vo\ra$ & $\hat\xi$ & $V_0$ \\
\hline
E1 & 0.20 & $4$ & 15.4545 & 17.1932 & 3384.8              & 6.50162 & $-1.9727\times 10^{-12}$ \\
E2 & 0.20 & $3$ & 39.6462 & 41.3674 & 47190.1             & 6.50162 & $-5.1269\times 10^{-16}$ \\
E3 & 0.20 & $2$ & 267.289 & 269.001 & 1.2977$\times 10^7$ & 6.50162 & $-1.6471\times 10^{-23}$ \\
E4 & 0.10 & $4$ & 378.003 & 380.424 & 3.6704$\times 10^7$ & 18.3894 & $-1.0294\times 10^{-24}$ \\
E5 & 0.15 & $3$ & 163.346 & 165.325 & 3.0125$\times 10^6$ & 10.0099 & $-1.7065\times 10^{-21}$ \\
E6 & 0.25 & $2$ & 76.0825 & 77.6175 & 311734.7            & 4.65218 & $-1.3231\times 10^{-18}$ \\
\hline
\end{tabular}
\captionof{table}{Benchmark examples for model $M_{3,15}$ where we have set $K_{\rm cs} = A_1 = 1$, $W_0 = -1$, $a_1 = 2\pi/N$ and $A_2 = A_3 = 0$.}
	\label{tab_min-Model-D}
\end{center}

The emergence of this novel LVS vacuum can be qualitatively understood as follows. The $T_1$-dependent non-perturbative $W$, in combination with $\mathcal{O}(\alpha'^3)$ effects, stabilises $\tau_1$ at `small' size and the CY volume exponentially large. The remaining flat direction in the 2-cycle volume moduli space, which we will parametrise as $t_*$, is also lifted since the non-diagonality of $D_1$ introduces a dependence of $V$ on $t_*$. Notice that the axions $\rho_2$ and $\rho_3$ can become massive only after including $T_2$- and $T_3$-dependent non-perturbative contributions to $W$. These terms would however be negligible for the stabilisation of the 2-cycle volume moduli since all our minima are located at $\tau_2=\tau_3 \gg \tau_1$, as can be seen from Tab. \ref{tab_min-Model-D1}.

It is important to stress here that in general the non-diagonality of a dP divisor modifies the scaling behaviour with the overall volume of the different contributions to $V$, so destroying the existence of an LVS minimum. However in our model this is not the case due to the presence of a symmetry of the moduli space under the exchange $t_2 \leftrightarrow t_3$, as can be seen from (\ref{M315data}) and (\ref{Topol}).

\begin{center}
\begin{tabular}{|c||c|c|c|c|} 
\hline
Example  & $\la\tau_1\ra$ & $\la\tau_2\ra$ & $\la\tau_3\ra$ & $\la\vo\ra$ \\
\hline
E1 & 6.04622 & 17.1932 & 17.1932 & 3384.8 \\
E2 & 5.92451 & 1708.3 & 1708.3 & 47190.1 \\
E3 & 5.86144 & 72358.4 & 72358.4 & 1.2977$\times10^7$ \\
E4 & 11.7229 & 144716.88 & 144716.88 & 3.6704$\times10^7$ \\
E5 & 7.83456 & 27328.40 & 27328.40 & 3.0125$\times10^6$ \\
E6 & 4.71267 & 6022.13 & 6022.13 & 311734.7 \\
\hline
\end{tabular}
\captionof{table}{Values of the 4-cycle moduli for the benchmark examples of model $M_{3,15}$ in Tab. \ref{tab_min-Model-D}.}
\label{tab_min-Model-D1}
\end{center}

In the large volume limit where $\vo\gg \hat\xi$ and $a_1\tau_1 \gg 1$, and after stabilising the axion $\rho_1$ such that $\cos(a_1 \la\rho_1\ra + \theta_0 - \phi_1)=-1$, the scalar potential of this model derived from (\ref{MasterF}) can be very well approximated as:
\be
V = \frac{e^{K_{\rm cs}}}{2 s} \left(\frac{4 |A_1|^2 a_1^2}{\vo^2} \,h(t_i)\, e^{- 2 a_1\tau_1} - \frac{4 |W_0| |A_1| a_1}{\vo^2}\,\tau_1\, e^{- a_1 \tau_1} + \frac{3 \hat\xi |W_0|^2}{4\vo^3}\right), 
\label{LargeVS}
\ee
where:
\be
h(t_i) = - \vo \left( \sum_{k=1}^3 k_{11k} t_k\right)  + \tau_1^2 = 2 \vo \left(t_2 + t_3 - 2\,t_1\right) +\tau_1^2\,.
\ee
Thanks to the symmetry of the moduli space under the exchange $t_2 \leftrightarrow t_3$, if we now write $t_3 = t_2 + t_*$, the $h(t_i)$ function takes the simple form:
\be
h(t_i) = 2 \vo \sqrt{2\tau_1 + t_*^2} + \tau_1^2\,.
\ee 
The potential (\ref{LargeVS}) therefore depends on only 3 variables: $\vo$, $\tau_1$ and $t_*$. The dependence on $t_*$ is very simple, signaling that there is a minimum at $t_*=0$, which implies $t_2=t_3$ and $\tau_2=\tau_3$ from (\ref{Topol}). Notice that this minimum lies well inside the K\"ahler cone since $t_*=0$ does not correspond to any boundary of the moduli space. More interestingly, for $t_*=0$, the potential (\ref{LargeVS}) takes the standard LVS form with $h(t_i)\simeq 2\sqrt{2}\,\vo \sqrt{\tau_1}$ for $\vo \gg \tau_1^{3/2}$: 
\be
V = \frac{e^{K_{\rm cs}}}{2 s} \left(8\sqrt{2} |A_1|^2 a_1^2 \,\sqrt{\tau_1}\, \frac{e^{- 2 a_1\tau_1}}{\vo} - 4 |W_0| |A_1| a_1\,\tau_1\, \frac{e^{- a_1 \tau_1}}{\vo^2} +\frac{3 \hat\xi |W_0|^2}{4\vo^3}\right).
\label{LargeVS2}
\ee
This potential has an LVS AdS minimum located at:
\be
\la\vo\ra\simeq \frac{|W_0| \sqrt{\la\tau_1\ra}}{4 \sqrt{2} a_1 A_1}\,e^{a_1\la\tau_1\ra}\qquad\text{and}\qquad \la\tau_1\ra \simeq \left(\frac{3\hat\xi}{\sqrt{2}}\right)^{2/3}\,.
\ee
These relations correctly reproduce the scaling behaviour of the numerical solutions presented in Tabs. \ref{tab_min-Model-D} and \ref{tab_min-Model-D1}. Let us stress that this is the first example of a CY threefold which admits LVS vacua even without the presence of a diagonal dP divisor, implying that LVS vacua in the string landscape occur more frequently than previously thought.

\subsubsection{New hybrid vacua}
\label{Hybrid}

\begin{figure}[H]
\begin{center}
\hspace*{-0.5cm} \includegraphics[width=14.0cm,height=8.0cm]{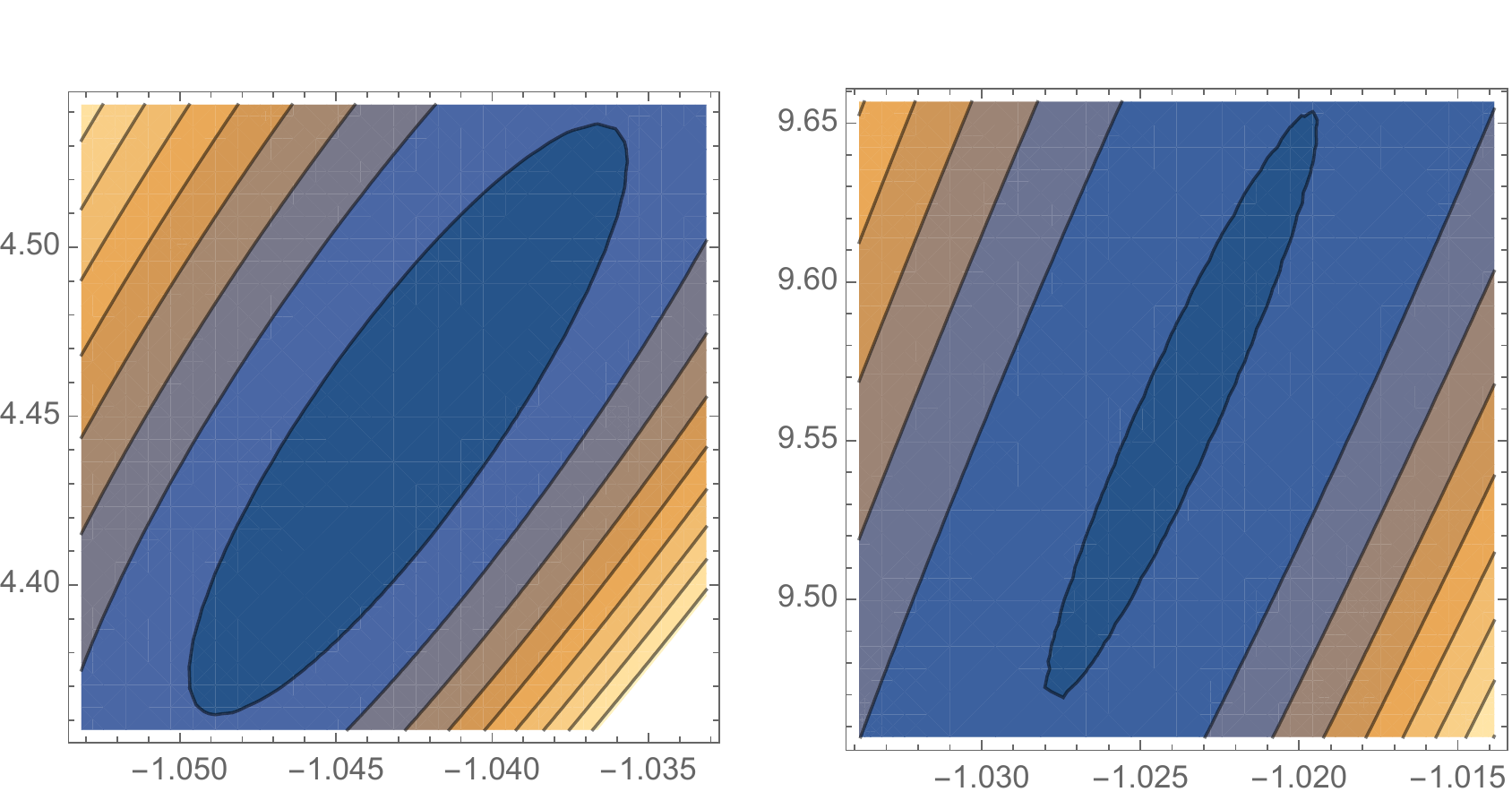}
\end{center}
\caption{Contour plot of the scalar potential of the 2 K\"ahler moduli $(t_1, t_2)$ for examples E1 (left) and E5 (right) of Tab. \ref{tab_min-Model-B}.}
\label{fig_Bnew}
\end{figure}

In this section we shall study if numerical moduli stabilisation in terms of 2-cycle moduli can reveal the existence of new LVS vacua in `hard' $h^{1,1}=2$ models where $\vo$ does not admit a simple form in terms of 4-cycle moduli. We will find that the absence of a diagonal dP divisor combined with the simplicity of this model which has only 2 K\"ahler moduli prevents the existence of an LVS-like vacuum. We will discover instead new vacua which we term `hybrid' since they share some features with standard LVS vacua and others with typical KKLT models. We shall illustrate our claim by focusing on model $M_{2,20}$ in Tab. \ref{tab_cydata-h11eq2} in App. \ref{AppE}. The CY threefold of model $M_{2,20}$ corresponds to the polytope ID $\#23$ in the CY database of \cite{Altman:2014bfa} and it is defined by the following toric data:
\begin{center}
\begin{tabular}{|c|cccccc|}
\hline
    CY & $x_1$  & $x_2$  & $x_3$  & $x_4$  & $x_5$ & $x_6$      \\
    \hline
7 & 0 & 1  & 1 & 2 & 1 & 2  \\
3 & 1 & 0  & 0 & 1 & 0 & 1  \\
\hline
 & dP$_1$  & SD$_1$ & SD$_1$ &  SD$_2$ & SD$_1$ & SD$_2$  \\
\hline
\end{tabular}
\end{center}
The Hodge numbers are $(h^{2,1}, h^{1,1}) = (95, 2)$, the Euler number is $\chi=-186$ and the SR ideal is ${\rm SR} =  \{x_1 \, x_4 \, x_6, \,  \, x_2 \, x_3 \, x_5\}$. The analysis of the divisor topologies shows that they are represented by the following Hodge diamonds:
\bea
& & {\rm dP}_1 \equiv \begin{tabular}{ccccc}
    & & 1 & & \\
   & 0 & & 0 & \\
  0 & & 2 & & 0 \\
   & 0 & & 0 & \\
    & & 1 & & \\
  \end{tabular}, \quad {\rm SD}_1 \equiv \begin{tabular}{ccccc}
    & & 1 & & \\
   & 0 & & 0 & \\
  2 & & 30 & & 2 \\
   & 0 & & 0 & \\
    & & 1 & & \\
  \end{tabular}, \quad {\rm SD}_2 \equiv \begin{tabular}{ccccc}
    & & 1 & & \\
   & 0 & & 0 & \\
  7 & & 66 & & 7 \\
   & 0 & & 0 & \\
    & & 1 & & \\
  \end{tabular}\,. \nn
\eea
As can be seen from Tab. \ref{tab_cydata-h11eq2}, the intersection numbers and the K\"ahler cone in the basis of smooth divisors $D_1=\{0,1\}={\rm dP}_1$ and $D_2 =\{2,1\} = {\rm SD}_2$ are:
\be
k_{111} = 8\,,\quad k_{112} = -2\,, \quad  k_{122} = 0\,, \quad k_{222} = 14\,, \qquad \text{K\"ahler cone:} \quad t_1 < 0\,, \, \, t_1 + t_2 > 0\,, \nn
\ee
which shows clearly that the dP$_1$ divisor $D_1$ is non-diagonal. The CY volume and the 4-cycle moduli become:
\be
\vo = \frac43\,t_1^3 - t_1^2\, t_2  + \frac73\,t_2^3\,, \qquad \tau_1 = 4\, t_1^2 - 2\, t_1\, t_2\,, \qquad \tau_2 = 7 \,t_2^2- t_1^2 \,. 
\label{Invert}
\ee
The potential can be obtained from our master formula (\ref{MasterF}) by setting $h^{1,1} = 2$, $n = 1$ and $\hat\xi\neq 0$. In Tab. \ref{tab_min-Model-B} we present the results of our numerical minimisation with respect to 2-cycle moduli for different choices of the microscopic parameters, while Fig. \ref{fig_Bnew} shows a contour plot of the potential for 2 particular examples (E1 and E5). Notice that the effective field theory is still under control even if $|\la t_1\ra|\sim\mathcal{O}(1)$ since all examples in Tab. \ref{tab_min-Model-B} satisfy the condition (\ref{cond1}) which guarantees that stringy corrections can be neglected.

\begin{center}
\begin{tabular}{|c||c|c|c|c|c|c|c|c|c||} 
\hline
Example & $g_s$ & $W_0$ & $a_1$ & $\la t_1\ra$ & $\la t_2\ra$ & $\la\vo\ra$ & $\hat\xi$ & $V_0\,(\times 10^{11})$ \\
\hline
E1 & 0.13 & $-$0.1 & $\pi/4$ & $-$1.04315 & 4.44218 & 198.187 & 9.61509 & $-$22.4614 \\
E2 & 0.10 & $-$0.2 & $\pi/5$ & $-$1.05354 & 5.87528 & 465.138 & 14.2518 & $-$5.15813 \\
E3 & 0.09 & $-$0.8 & $\pi/6$ & $-$1.00075 & 7.18459 & 856.799 & 16.6919 & $-$9.78060 \\
E4 & 0.08 & $-$1.0 & $\pi/7$ & $-$1.07603 & 7.48456 & 967.979 & 19.9174 & $-$12.2713 \\
E5 & 0.07 & $-$1.8 & $\pi/8$ & $-$1.02386 & 9.55669 & 2025.12 & 24.3344 & $-$2.76468 \\
\hline
\end{tabular}
\captionof{table}{Benchmark examples for model $M_{2,20}$ where we have set $K_{\rm cs} = 1$, $A_1=10$ and $A_2 = 0$.}
\label{tab_min-Model-B}
\end{center}

We stress that we considered only a $T_1$-dependent non-perturbative $W$ since $A_2=0$, and so the K\"ahler moduli are fixed by balancing the leading $\alpha'$-correction to $K$ against the superpotential generated by gaugino condensation in a hidden gauge group with rank $N$. This is the same stabilisation mechanism  used in LVS models which are characterised by an exponentially large volume $\la \vo\ra \sim e^{1/(g_s N)}$ that increases when either $g_s$ or $N$ decreases. However, as can be seen from Tab. \ref{tab_min-Model-B} where we limit ourselves to $N \leq 16$, in our new vacua $\la\vo\ra$ increases when $g_s$ decreases but it reduces when $N$ goes to lower values. Hence these new vacua are not really LVS-like. This difference can be traced back to the absence of a diagonal dP divisor together with the fact that this model has only 2 K\"ahler moduli while the CY threefold discussed in Sec. \ref{NewLVS2} had $h^{1,1}=3$.

The behaviour of these new vacua can be understood analytically as follows. Tab. \ref{tab_min-Model-B} shows that, in all examples, the 2-cycle volume moduli $t_1$ and $t_2$ are fixed at $|t_1| \simeq 1$ and $t_2^2 \gg t_1^2$. In this limit of the K\"ahler cone, the expressions for $\vo$ and $\tau_2$ in (\ref{Invert}) simplify to $\tau_2 \simeq 7\, t_2^2$  and $\vo \simeq \frac73\, t_2^3 \simeq \frac{1}{3\sqrt{7}}\,\tau_2^{3/2}$. Moreover, as can be seen from Tab. \ref{tab_min-Model-B}, all solutions are located at $\vo \gg \hat\xi$ and $a_1\tau_1\gg 1$, and so our master formula (\ref{MasterF}) can be very well approximated by (after $\rho_1$ minimisation):
\be
V \simeq  - 4 \,a_1^2\, |A_1|^2  f(\tau_1,\vo) \,\frac{e^{-2 a_1 \tau_1}}{\vo} -  4 \, |W_0| \, |A_1|\,a_1\, \tau_1\,\frac{e^{- a_1 \tau_1}}{\vo^2}+\frac{3 \, \hat\xi\,|W_0|^2}{4\vo^3}\,,
\label{Vsimpl}
\ee
where:
\be
f(\tau_1,\vo) = \left(k_{111} t_1 - k_{112} t_2\right) \simeq -2 \,\sqrt{\left(\frac37 \,\vo\right)^{2/3} + 4 \tau_1}\,.
\label{fvo}
\ee
Notice that, similarly to LVS models, the negative sign in (\ref{fvo}) is crucial to find a minimum. However the potential (\ref{Vsimpl}) does not give rise to a minimum at exponentially large volume. In fact, if one takes the limit $\vo \gg \tau_1^{3/2}$, the function in (\ref{fvo}) simplifies to $f(\tau_1, \vo)\simeq -2 \,\left(\frac37 \,\vo\right)^{1/3}$ and the potential (\ref{Vsimpl}) becomes schematically:
\be
V = c_1 \,\frac{e^{-2 a_1 \tau_1}}{\vo^{2/3}} -  c_2\, \tau_1\,\frac{e^{- a_1 \tau_1}}{\vo^2}+\frac{c_3}{\vo^3}\qquad \text{with}\qquad c_{1,2,3}>0\,.
\ee
Extremising this potential with respect to $\tau_1$ yields:
\be
\frac{\partial V}{\partial \tau_1}=0\qquad \Leftrightarrow \qquad \vo = \left(\frac{c_2}{2 c_1}\right)^{3/4} \tau_1^{3/4}\,e^{\tfrac34 a_1\tau_1}\,,
\ee
which appears to indicate the presence of a solution at exponentially large volume. However using this expression to integrate out $\tau_1$, the scalar potential reduces to:
\be
V \simeq \frac{- c_4 \left(\ln\vo\right)^2 + c_3 \,\vo^{1/3}}{\vo^{10/3}}\qquad \text{with}\qquad c_4 = \frac{4 c_2^2}{9 a_1^2 c_1}>0\,.
\label{VnoLVS}
\ee
The behaviour of this potential at large volume is qualitatively different from standard LVS models where, after integrating out $\tau_1$, one has:
\be
V_{\rm LVS} \simeq \frac{- \lambda_1 \left(\ln\vo\right)^{3/2} + \lambda_2}{\vo^3}\qquad \text{with}\qquad \lambda_{1,2}>0\,.
\label{VyesLVS}
\ee
For $\vo \gg 1$, the LVS potential (\ref{VyesLVS}) is dominated by the logarithmic term proportional to $\lambda_1$, and so it goes to zero from negative values, while for $\vo \sim \mathcal{O}(1)$ it is dominated by the term proportional to $\lambda_2$, and so it is positive. Clearly the potential has to admit an AdS minimum at large volume. On the other hand, the potential (\ref{VnoLVS}) is not guaranteed to feature a minimum since the term proportional to $c_3$ dominates the potential for both $\vo \sim \mathcal{O}(1)$ and $\vo \gg 1$, implying that at large volume it goes to zero from positive values. In fact there is a window at intermediate volume values, i.e. for $\mathcal{O}(1) \lesssim \vo \lesssim \mathcal{O}(100)$, where the 2 terms in (\ref{VnoLVS}) can compete. This reveals the existence of a minimum, which is however in a strong string coupling regime where the effective field theory is out of control, and only a maximum at exponentially large volume. However, given that the axion $\rho_1$ has been kept fixed at its minimum, this would-be maximum is actually a saddle point.

The presence of a saddle point at exponentially large volume can be explicitly seen by taking the first and second derivatives of (\ref{VnoLVS}) with respect to $\vo$ which read as follows (for $x\equiv \ln \vo$):
\be
\frac{\partial V}{\partial \vo}=0\quad \Leftrightarrow \quad \vo^{1/3} = \frac{10 c_4}{9 c_3}\, x^2 \left(1-\frac{3}{5 x}\right)\quad\Rightarrow\quad \vo\,\frac{\partial^2 V}{\partial \vo^2} = -\frac{10 c_4}{3}\, x^2 \left(1 - \frac{33}{5 x} + \frac{9}{5 x^2}\right). 
\label{above}
\ee
For $x\gg 1$, the second derivative is clearly negative, signaling the existence of a saddle point. The second derivative can actually become positive, so giving rise to a minimum, for $x\lesssim 6.3$. In order to trust the initial approximation $f(\tau_1, \vo)\simeq -2 \,\left(\frac37 \,\vo\right)^{1/3}$ for the function in (\ref{fvo}), one needs to have at least $x\gtrsim 5.3$ for $\tau_1\gtrsim 1$ so that $4\tau_1 \left(\frac37 \,\vo\right)^{-2/3}\lesssim 0.2$. Minima with $5.3\lesssim x < 6$ would be AdS, $x=6$ would give Minkowski and $6<x\lesssim 6.3$ would yield dS. However one can check that none of these vacua can be trusted since they would lie at $g_s>1$, in a regime where perturbation theory would break down. This can be easily seen by using the minimisation equation in (\ref{above}) as an expression for $c_3$ with the volume fixed in the regime $5.3\lesssim x \lesssim 6.3$, and then using this result to find the value of $g_s$ knowing that $c_3$ can be also expressed as $c_3 = 3\xi\,W_0^2/(4 g_s^{3/2})$.

This discussion implies that there is no minimum in the region of moduli space where the function in (\ref{fvo}) can be approximated as $f(\tau_1, \vo)\simeq -2 \,\left(\frac37 \,\vo\right)^{1/3}$. In fact, defining:
\be
R\equiv \left(\frac37\right)^{2/3} \frac{\la\vo\ra^{2/3}}{4\la\tau_1\ra}\,,
\label{Rdef}
\ee
the two terms in (\ref{fvo}) are always of the same order of magnitude since $0.5 \lesssim R \lesssim 1$ for all vacua listed in Tab. \ref{tab_min-Model-B}. The potential (\ref{Vsimpl}) can then be minimised analytically giving (for $a_1\la\tau_1\ra \gg 1$):
\be
\la\vo\ra\simeq \frac{d_1}{a_1} \sqrt{\la\tau_1\ra}\,e^{a_1\la\tau_1\ra}\qquad\text{and}\qquad \la\tau_1\ra\simeq \,\frac{d_2}{g_s}\,,
\label{MinPos}
\ee
with:
\be
d_1  = \frac{|W_0|}{8 |A_1| \sqrt{1+R}}\qquad\text{and}\qquad d_2  = \left(3\xi\right)^{2/3} \left(1+R\right)^{1/3}\,.
\ee
The relations (\ref{MinPos}) resemble those of standard LVS AdS vacua but the condition (\ref{Rdef}) with $R\sim \mathcal{O}(1)$ now implies also (setting $a_1=2\pi/N$):
\be
N\, g_s \simeq \frac{d_2}{d_3} \qquad\text{with}\qquad d_3 = \frac{1}{2\pi}\ln \left[\left(\frac{d_4}{d_1}\right)\ln \left(\frac{d_4}{d_1}\right)\right] \qquad\text{and}\qquad d_4 = \frac{56}{3}\,R^{3/2}\,.
\ee
Interestingly, for all examples in Tab. \ref{tab_min-Model-B}, $d_2 \simeq d_3 \simeq \mathcal{O}(1)$, and so $N\,g_s\simeq \mathcal{O}(1)$.
Thus the value of the overall volume at the AdS minimum is given by:
\be
\la\vo\ra \simeq d_4 \,\la\tau_1\ra^{3/2} \simeq \left(d_4 \,d_2^{3/2}\right) \frac{1}{g_s^{3/2}} \simeq \left(d_4 \,d_2^{3/2}\right) N^{3/2}\,, 
\label{voMin}
\ee
which reproduces the behaviour of the volume in Tab. \ref{tab_min-Model-B}, since $\la\vo\ra$ increases when either $g_s$ decreases or $N$ increases. Moreover (\ref{voMin}) implies that the combination $r\equiv \la\vo\ra (g_s/R)^{3/2}$ should be more or less constant, and this is confirmed by all examples in Tab. \ref{tab_min-Model-B} which feature $r\simeq \mathcal{O}(40)$. This analytical estimate is useful also to perform our numerical study since it provides reasonable initial conditions to easily find convergent solutions.

It is for these reasons that we term these new vacua `hybrid': they clearly have some similarities and also some differences with known stabilisation mechanisms which can be summarised as follows:
\bi
\item \textbf{KKLT}: both cases admit an AdS vacuum where $\la\vo\ra \propto N^{3/2}$ even if, contrary to KKLT, in our new vacua supersymmetry is broken, $\alpha'$ effects play a crucial r\^ole and $|W_0|$ does not need to be tuned exponentially small;

\item \textbf{LVS}: both cases feature a non-supersymmetric AdS vacuum where non-perturbative effects compete with $\alpha'$ corrections for natural values of $|W_0|$ even if, contrary to LVS, in our new vacua the volume in string units is not exponentially large;

\item \textbf{$\alpha'$ uplift}: both cases have a minimum which breaks supersymmetry via balancing non-perturbative against $\alpha'$ contributions without tuning $|W_0|$ even if, contrary to $\alpha'$ uplift, our new vacua are AdS and require non-perturbative effects just for 1 modulus.
\ei

\section{Conclusions}
\label{Conclusions}

In this article we have presented a new systematic approach to type IIB moduli stabilisation which is based on fixing the K\"ahler moduli through the 2-cycle volume moduli as opposed to the standard approach which uses the 4-cycle volume moduli.

With the help of numerical techniques, we have been able to reproduce all known approaches to type IIB K\"ahler moduli stabilisation and to identify new classes of models that could not be determined with previous methods. In particular we discovered the first examples in the literature of LVS vacua for CY threefolds which do not admit a diagonal dP divisor. This implies that the presence of LVS vacua in the string landscape is more generic than previously thought. An interesting future line of investigation would be to perform a more systematic analysis of the frequency of LVS in type IIB flux compactifications.

Moreover our innovative approach to K\"ahler moduli stabilisation allowed us also to reveal a new class of hybrid models where the volume is stabilised at values large enough to be of phenomenological and cosmological interest, as well as to guarantee control over the effective field theory approximation, but not exponentially large as in standard LVS models. More work in this direction is certainly needed, both to explore the physical implications of this new class of models and also to obtain better computational control over the effective field theory.

In order to consider concrete models, we have been systematic in our approach and started by covering all known models constructed from hypersurfaces in toric varieties by Kreuzer and Skarke with Hodge numbers $h^{1,1}=1,2,3$. We have classified them according to whether they are of the LVS type: standard Swiss cheese LVS models, K3 fibrations with a diagonal dP divisor, strong Swiss cheese-like examples and structureless LVS CY models which can still lead to stabilised vacua with exponentially large internal volume.

The underlying message for analysing the dataset with $1 \leq h^{1,1} \leq 3$ is that, while all examples with $h^{1,1} =1$ can be studied via the conventional approach based on 4-cycle moduli, only 72\% of the models with $h^{1,1} = 2$ and 50\% of the models with $h^{1,1}=3$ can be analysed with this standard approach. Thus the new strategy described in our paper to find stable vacua by working in terms of 2-cycle volume moduli is essential for achieving  full moduli stabilisation with supersymmetry breaking for $h^{1,1}=3$, and indeed our approach seems to be the only way to proceed for larger $h^{1,1}$.

This article can then be considered as only the first step towards the more systematic aim of performing full moduli stabilisation with an arbitrarily large number of K\"ahler moduli. With the analytic and numerical techniques developed in this article, we hope in the future to approach concrete models with $h^{1,1}\geq 4$, possibly even in the regime of large $h^{1,1}\sim {\mathcal{O}} (10^2-10^3)$ where we may be able to use a large Hodge number approximation. 

Interesting directions for future work are the inclusion in our master formula of string loop corrections and higher order $\alpha'$ effects, as well as a detailed exploration of the axion landscape for cases with large $h^{1,1}$. The presence of many axions allows for a potentially large landscape inside the actual string landscape, with the added value that extrema should be computable within the effective field theory as proposed in \cite{Bachlechner:2017zpb, Bachlechner:2017hsj, Bachlechner:2019vcb}. We hope to come back and address these questions in the future.

\section*{Acknowledgments}

We would like to thank Francesco Muia, Andreas Schachner and Roberto Valandro for useful conversations and ICTP for hospitality at different stages of this project. We especially thank Nicole Edmea Bollan for collaborations at an early stage of this project.

\appendix

\section{Known potentials from our master formula}
\label{AppA}

In this appendix we will show how our master formula (\ref{MasterF}) reduces to different known scalar potentials by just choosing 3 parameters: the Hodge number $h^{1,1}$, the number $n$ of non-perturbative contributions to the superpotential, and CY Euler-number $\hat\xi$. Some of these models are collected in Tab. \ref{tab_known-models}.

\subsection{1-modulus KKLT}

For reproducing the standard KKLT potential \cite{Kachru:2003aw}, we need to consider $h^{1,1} = n = 1$ and $\hat\xi = 0$. In this case the 3 contributions to the general potential given in (\ref{MasterF}) become:
\bea
V_{\mathcal{O}(\alpha'^3)} &=& 0, \quad V_{\rm np1} =  4 \, e^{{\cal K}}\, |W_0| \, |A_1|\,a_1\, \tau_1\,e^{- a_1\, \tau_1}\, \cos(a_1 \rho_1 + \theta_0 - \phi_1),  \nn \\
V_{\rm np2} &=& 4\, e^{{\cal K}}\, |A_1|^2 \, e^{- 2 a_1 \tau_1} \left( - \vo\, a_1^2\,k_{111}\,t_1 +  (a_1\, \tau_1)^2 +  a_1 \, \tau_1\right),
\eea
where:
\be
\vo = \frac16\,k_{111}\, t_1^3\,, \qquad \tau_1 = \frac12\,k_{111}\, t_1^2\,, \qquad e^{{\cal K}} = \frac{e^{K_{\rm cs}}}{2\, s \, \vo^2}\,.
\ee
This leads to the standard KKLT scalar potential which admits a supersymmetric AdS vacuum:
\be
V_{\rm KKLT} = \frac{9\, e^{K_{\rm cs}}\, a_1\, k_{111}\, |A_1|}{s\, \tau_1^2}\, e^{-a_1\, \tau_1} \left[ |W_0| \, \cos(a_1 \rho_1 + \theta_0 - \phi_1) + \frac{|A_1|}{3} \, e^{-a_1\, \tau_1} \left(a_1\, \tau_1 +3 \right)\right]. \nn
\ee

\subsection{2-moduli KKLT}

For reproducing the potential of KKLT models with 2 K\"ahler moduli \cite{Denef:2004dm, BlancoPillado:2006he}, we need to consider $h^{1,1} = n = 2$ and $\hat\xi = 0$. In this case the 3 contributions in (\ref{MasterF}) become:
\bea
V_{\mathcal{O}(\alpha'^3)} &=&  0\,, \qquad V_{\rm np1} = \frac{e^{K_{\rm cs}}}{2 s \vo^2}\,\sum_{i=1}^2 4 a_i \tau_i |W_0| |A_i|\, \cos(a_i \rho_i + \theta_0 - \phi_i)\, e^{- a_i \tau_i}\,, \nn \\
V_{\rm np2} &=& \frac{e^{K_{\rm cs}}}{2 s \vo^2} \left[ \sum_{i=1}^2  4 a_i^2 |A_i|^2 \left(- \vo\, k_{iii} t_i + \tau_i^2 + \frac{\tau_i}{a_i} \right)\, e^{-2 a_i \tau_i}\right. \label{KKLT2} \\
&+& \left. 8 a_1 a_2 |A_1| |A_2| \, e^{-(a_1 \tau_1 + a_2 \tau_2)}\cos(a_1 \rho_1 - a_2 \rho_2 - \phi_1 +\phi_2) \left(\tau_1 \tau_2 + \frac{a_1\tau_1+a_2\tau_2}{2 a_1 a_2} \right) \right], \nn
\eea
where, similarly to the $\mathbb{C}{\rm P}^4[{1,1,1,6,9}]$ model studied in \cite{Denef:2004dm}, we focused on a CY example with only $k_{111}\neq 0$ and $k_{222}\neq 0$. Hence the 4-cycle moduli and the overall volume read:
\bea
\tau_1 &=& \frac12\,k_{111} t_1^2 \quad \Rightarrow \quad t_1 = - \sqrt{\frac{2 \, \tau_1}{k_{111}}} \qquad\text{and}\qquad k_{111} t_1 = - \sqrt{2 k_{111} \tau_1}\,, \nn \\
\tau_2 &=& \frac12\,k_{222} t_2^2 \quad \Rightarrow \quad t_2 = + \sqrt{\frac{2 \, \tau_2}{k_{222}}}\qquad\text{and}\qquad k_{222} t_2 = + \sqrt{2 k_{222} \tau_2} \,, \nn \\
\vo &=& \frac16\left(k_{111}\,t_1^3 + k_{222}\,t_2^3\right)  =  \frac{\sqrt{2}}{3\sqrt{k_{222}}}\left( \tau_2^{3/2} - \sqrt{\frac{k_{222}}{k_{111}}}\,\tau_1^{3/2}\right), \label{TopKKLT}
\eea
where the minus sign in the relation between $t_1$ and $\tau_1$ is due to the fact that $\tau_1$ is the volume of a diagonal dP divisor whose K\"ahler cone condition is $t_1<0$. Thus the potential (\ref{KKLT2}) reduces to the standard 2-moduli KKLT form mentioned in \cite{BlancoPillado:2006he}:
\bea
V &=& \frac{e^{K_{\rm cs}}}{2 s \vo^2}\left[ \sum_{i=1}^2 4 a_i \tau_i |W_0| |A_i|\, \cos(a_i \rho_i + \theta_0 - \phi_i)\, e^{- a_i \tau_i}\right. \nn \\
&+& \sum_{\substack{i=1 \\ i\neq j}}^2 \frac43\,|A_i|^2 a_i^2 \tau_i^2 \, e^{-2 a_i \tau_i} \left(1 + 2 \sqrt{\frac{k_{iii}}{k_{jjj}}}\,\left(\frac{\tau_j}{\tau_i}\right)^{3/2} + \frac{3}{a_i \tau_i} \right) \nn \\
&+& \left. 8 a_1 a_2 |A_1| |A_2| \, e^{-(a_1 \tau_1 + a_2 \tau_2)}\cos(a_1 \rho_1 - a_2 \rho_2 - \phi_1 +\phi_2) \left(\tau_1 \tau_2 + \frac{a_1\tau_1+a_2\tau_2}{2 a_1 a_2} \right) \right]. \nn
\eea

\subsection{2-moduli Swiss cheese LVS}

For reproducing the standard LVS potential \cite{Balasubramanian:2005zx,Cicoli:2012vw, Cicoli:2013mpa, Cicoli:2013cha}, let us consider $h^{1,1}=2$, $n=1$ and $\hat\xi>0$. In this case the 3 contributions to the general potential given in (\ref{MasterF}) become:
\bea
V_{\mathcal{O}(\alpha'^3)} &=&  e^{{\cal K}} \, \frac{3 \, \hat\xi \, (\vo^2 + 7\, \hat\xi \,\vo +\hat\xi^2)}{(\vo-\hat\xi) (2\vo +\hat\xi)^2}\, \,|W_0|^2,\\
V_{\rm np1} &=& 2\, e^{{\cal K}} \, |W_0| \, |A_1|\, e^{- a_1 \tau_1}\, \cos(a_1 \rho_1 + \theta_0 - \phi_1) \nn \\
& & \times \biggl[\frac{(4\,\vo^2  + \vo \, \hat\xi +4\, \hat\xi^2)}{(\vo - \hat\xi) (2\, \vo + \hat\xi)}\, (a_1 \tau_1) +\frac{3 \, \hat{\xi } \, (\vo^2 + 7\, \hat\xi \,\vo +\hat\xi^2)}{(\vo-\hat\xi)\,(2\vo +\hat\xi)^2}\biggr],  \nn \\
V_{\rm np2} &=& 4\,e^{{\cal K}}\, |A_1|^2 \, e^{- 2 a_1 \tau_1} \biggl[ - \left(\vo+\frac{\hat\xi}{2}\right) \,a_1^2\, k_{111}\,t_1  + \frac{4\vo - \hat\xi}{4(\vo - \hat{\xi})} (a_1 \tau_1)^2 \nn \\
& & + \frac{(4\,\vo^2  + \vo \, \hat\xi + 4\, \hat\xi^2)}{2 (\vo - \hat\xi) (2\, \vo+ \hat\xi)}\, (a_1 \tau_1) +\frac{3 \, \hat\xi\, (\vo^2 + 7\, \hat{\xi} \,\vo + \hat\xi^2)}{4 (\vo -\hat\xi) (2 \,\vo + \hat\xi)^2}\biggr]. \nn
\eea
Focusing on the large volume limit, the leading order contributions in all terms above give:
\be
V \simeq \frac{e^{K_{\rm cs}}}{2 s} \left[ \frac{3 \hat\xi |W_0|^2}{4 \vo^3} + \frac{4 a_1 \tau_1 |W_0| |A_1|}{\vo^2}\, e^{- a_1 \tau_1} \cos\left(a_1 \rho_1 + \theta_0 - \phi_1\right) \\
- \frac{4 a_1^2 |A_1|^2 k_{111} t_1}{\vo}\, e^{-2 a_1 \tau_1} \right]. 
\label{Vlvs}
\ee
In Swiss cheese LVS models with 2 K\"ahler moduli, the only non-zero intersection numbers are $k_{111}$ and $k_{222}$ and the relations between 2- and 4-cycle moduli look as in (\ref{TopKKLT}) where $\tau_1$ plays the r\^ole of the `small' modulus while $\tau_2$ corresponds to the `big' divisor. Hence (\ref{Vlvs}) takes the form:
\be
V \simeq \frac{e^{K_{\rm cs}}}{2 s} \left( \frac{\beta_{\alpha'}}{\vo^3} + \beta_{\rm np1}\,\frac{\tau_1}{\vo^2}\, e^{- a_1 \tau_1} \cos\left(a_1 \rho_1 + \theta_0 - \phi_1\right) \\
+ \beta_{\rm np2}\,\frac{ \sqrt{\tau_1}}{\vo}\, e^{-2 a_1 \tau_1} \right), 
\label{VlvsSimpl}
\ee
with:
\be
\beta_{\alpha'} = \frac{3 \hat\xi |W_0|^2}{4}\,, \qquad \beta_{\rm np1} = 4 a_1 |W_0| |A_1|\,, \qquad \beta_{\rm np2} = 4 a_1^2 |A_1|^2 \sqrt{2 k_{111}}\,. 
\ee
Notice that (\ref{VlvsSimpl}) matches the form of the potential of standard Swiss cheese LVS models with 2 K\"ahler moduli \cite{Balasubramanian:2005zx,Cicoli:2012vw, Cicoli:2013mpa, Cicoli:2013cha}.

\subsection{3-moduli Swiss cheese LVS}

The scalar potential of Swiss cheese LVS models with 3 K\"ahler moduli \cite{Conlon:2005jm, Cicoli:2017shd} can be reproduced by our master formula (\ref{MasterF}) by setting $h^{1,1}=3$, $n=2$ and $\hat\xi>0$, yielding:
\bea
\label{eq:Vgen-h113n1}
V_{\mathcal{O}(\alpha'^3)} &=&  e^{{\cal K}} \, \frac{3 \, \hat\xi (\vo^2 + 7 \hat\xi \vo +\hat\xi^2)}{(\vo-\hat\xi) (2\vo + \hat\xi)^2}\, |W_0|^2\,, \\
V_{\rm np1} &=& e^{{\cal K}}\, \sum_{i=1}^2 2 |W_0| |A_i|\, e^{- a_i \tau_i}\, \cos(a_i \rho_i + \theta_0 - \phi_i) \nn \\
&\times& \left[\frac{(4\vo^2 + \vo \hat\xi + 4 \hat\xi^2)}{(\vo - \hat\xi) (2\, \vo + \hat{\xi})}\, a_i\tau_i +\frac{3 \, \hat\xi (\vo^2 + 7 \hat\xi \vo +\hat\xi^2)}{(\vo-\hat\xi) (2\vo +\hat\xi)^2}\right]\, ,\nn \\
V_{\rm np2} &=& e^{{\cal K}} \, \sum_{i=1}^2   \sum_{j=1}^2 |A_i| |A_j| \, e^{- (a_i\tau_i + a_j\tau_j)} \, \cos(a_j \rho_j - a_i \rho_i -\phi_j + \phi_i)  \nn \\
&\times& \left[ -4 \left(\vo+\frac{\hat\xi}{2}\right) a_i a_j \left( \sum_{k=1}^3 k_{ijk} t_k\right)  + \frac{4 \vo - \hat\xi}{(\vo - \hat\xi)}\, a_i a_j \tau_i \tau_j\right. \nn \\
&+& \left.\frac{(4 \vo^2 + \vo \hat\xi + 4\hat\xi^2)}{(\vo -\hat\xi) (2 \vo + \hat\xi)}\, (a_i\tau_i +a_j\tau_j) +\frac{3 \, \hat\xi (\vo^2 + 7 \hat\xi \vo +\hat\xi^2)}{(\vo-\hat\xi) (2 \vo + \hat\xi)^2}\right]\,. \nn
\eea
In the large volume limit, this potential can very well be approximated as:
\bea
V &=& \frac{e^{K_{\rm cs}}}{2 s} \left[ \frac{3 \hat\xi |W_0|^2}{4\vo^3} + \sum_{i=1}^2 \frac{4 |W_0| |A_i| a_i}{\vo^2}\,\tau_i\, e^{- a_i \tau_i}\, \cos(a_i \rho_i + \theta_0 - \phi_i)\right. \label{3LVS} \\
&-& \left. \sum_{i=1}^2 \sum_{j=1}^2 \frac{4 |A_i| |A_j| a_i a_j}{\vo} \, e^{- (a_i\tau_i + a_j\tau_j)} \, \cos(a_j \rho_j - a_i \rho_i -\phi_j + \phi_i) \left(\sum_{k=1}^3 k_{ijk} t_k \right)\right]. \nn 
\eea
Given that we are interested in Swiss cheese CY models where the only non-vanishing intersection numbers are $k_{111}$, $k_{222}$ and $k_{333}$, we have:
\be
\sum_{k=1}^3 k_{iik}t_k = k_{iii} t_i = - \sqrt{2 \, k_{iii}\, \tau_i} \quad\text{for}\,\,i=1,2\qquad\text{and}\qquad \sum_{k=1}^3 k_{ijk}t_k = 0\quad\text{for}\,\,i\neq j\,. \nn
\ee
Hence (\ref{3LVS}) reduces to the potential of known 3-moduli Swiss cheese LVS models \cite{Conlon:2005jm, Cicoli:2017shd}:
\be
V = \frac{e^{K_{\rm cs}}}{2 s} \left[ \frac{\beta_{\alpha'}}{\vo^3} + \sum_{i=1}^2 \left(\beta_{{\rm np1},i}\,\frac{\tau_i}{\vo^2}\, e^{- a_i \tau_i}\, \cos(a_i \rho_i + \theta_0 - \phi_i) + \beta_{{\rm np2},i}\,\frac{\sqrt{\tau_i}}{\vo} \, e^{- 2 a_i\tau_i}\right) \right], \nn 
\ee
with:
\be
\beta_{\alpha'} = \frac{3 \hat\xi |W_0|^2}{4}\,, \qquad \beta_{{\rm np1},i} = 4 a_1 |W_0| |A_1|\,, \qquad \beta_{{\rm np2},i} = 4 a_1^2 |A_1|^2 \sqrt{2 k_{111}}\,. 
\ee

\subsection{3-moduli fibred LVS}

We now focus on 3-moduli fibred LVS models whose potential can be derived from our master formula (\ref{MasterF}) by setting $h^{1,1}=3$, $n=2$ and $\hat\xi>0$. Hence its form is the same as in (\ref{eq:Vgen-h113n1}) but now $\sum_{k=1}^3 k_{ijk} t_k$ is different since the underlying CY threefold has a distinct topological structure. In fact, in this case the CY features a K3 or $T^4$-fibration over a $\mathbb{P}^1$ base together with a diagonal dP divisor (for explicit CY threefolds with this topology see \cite{Cicoli:2011qg, Cicoli:2011it, Cicoli:2016xae, Cicoli:2017axo}). Via an appropriate choice of basis, the only non-zero intersection numbers can be chosen to be $k_{111}$ and $k_{233}$, signaling that $D_1$ is a diagonal dP divisor, $D_2$ is a K3 or $T^4$ fibre and $D_3$ contains the $\mathbb{P}^1$ base of the fibration. Thus we obtain:
\bea
\sum_{k=1}^3 k_{11k} t_k &=& -\, \sqrt{2 k_{111} \tau_1}\,, \qquad\text{and}\qquad \sum_{k=1}^3 k_{12k} t_k = \sum_{k=1}^3 k_{22k} t_k = 0\,, \nn \\
\vo &=& \frac16\,k_{111}\, t_1^3 + \frac12\, k_{233}\, t_2\, t_3^2 = \frac{\tau_3\, \sqrt{\tau_2}}{\sqrt{2} \, \sqrt{k_{233}}} -  \frac{\sqrt{2}\, \tau_1^{3/2}}{3\, \sqrt{k_{111}}}\,. 
\eea
By substituting these expression in (\ref{eq:Vgen-h113n1}) we can easily read off the potential of fibred LVS models with 3 K\"ahler moduli. Interestingly, in \cite{Cicoli:2008va} it has been shown that this potential cannot give rise to any LVS vacuum in a regime where the effective field theory is under control. In order to achieve this goal, one has to consider additional perturbative $\alpha'$ or $g_s$ corrections to the potential which are however not captured by our master formula (\ref{MasterF}).

\section{Lipschitz optimisation algorithm}
\label{AppC}

Exploring string theory models and related mathematical data vis-a-vis observations calls for new approaches to moduli stabilisation. Central to the stabilisation exercise is the need to minimise supergravity potential, a function of moduli fields that arises upon string compactification. The classical approach were via the selection of CY base geometry by hand and analytical minimisation of the one- or few-moduli potentials. In order to go beyond this, numerical automisation is required for selecting geometries and finding the positions of corresponding minima positions in moduli space. Here we describe a global optimisation algorithm, Lipschitz optimisation (LIPO), used for numerically stabilising the moduli fields of the example geometries addressed in this paper~\footnote{There are many optimisation algorithms in the literature. The application of these for CY selections and for finding corresponding minima in moduli space is an interesting research direction to pursue further.}.  

Lipschitz optimisation falls within the class of deterministic global optimisations (see for instance \cite{globop}). The task is to find the absolutely best set of parameters for achieving a mathematically-formulated objective. In~\cite{lipo} the LIPO algorithm for finding $x \in {\cal R}^d$ which globally maximises a function $f(x)$ was introduced. The DLIB library (see \url{dlib.net}) implements LIPO algorithm and improves it with local trust tests at the global maximum point. The basic principle for the Lipschitz optimisation is as follows.

\begin{figure}[!t]
\begin{center}
 \includegraphics[width=0.3\textwidth]{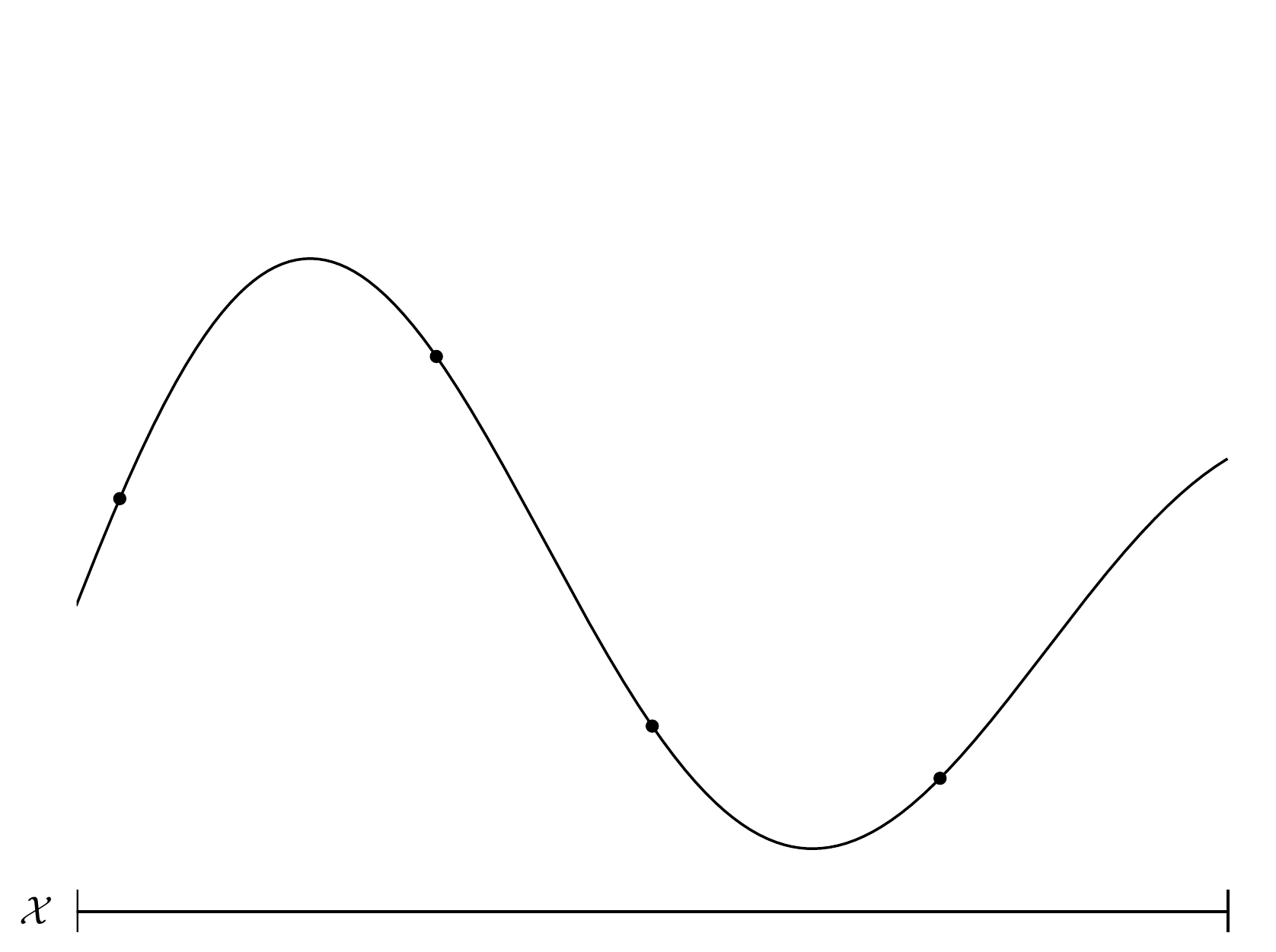}  
 \includegraphics[width=0.3\textwidth]{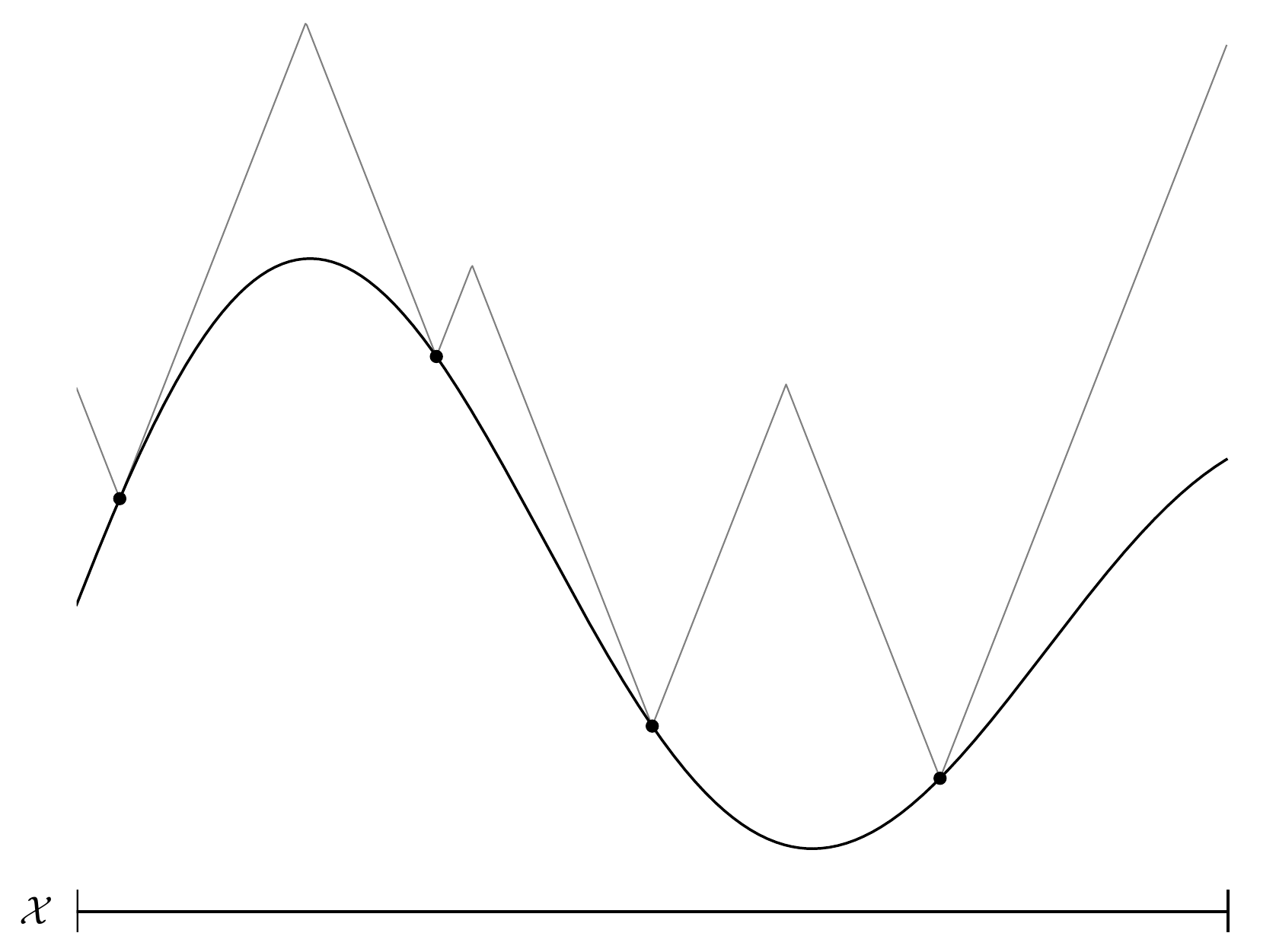}
 \includegraphics[width=0.3\textwidth]{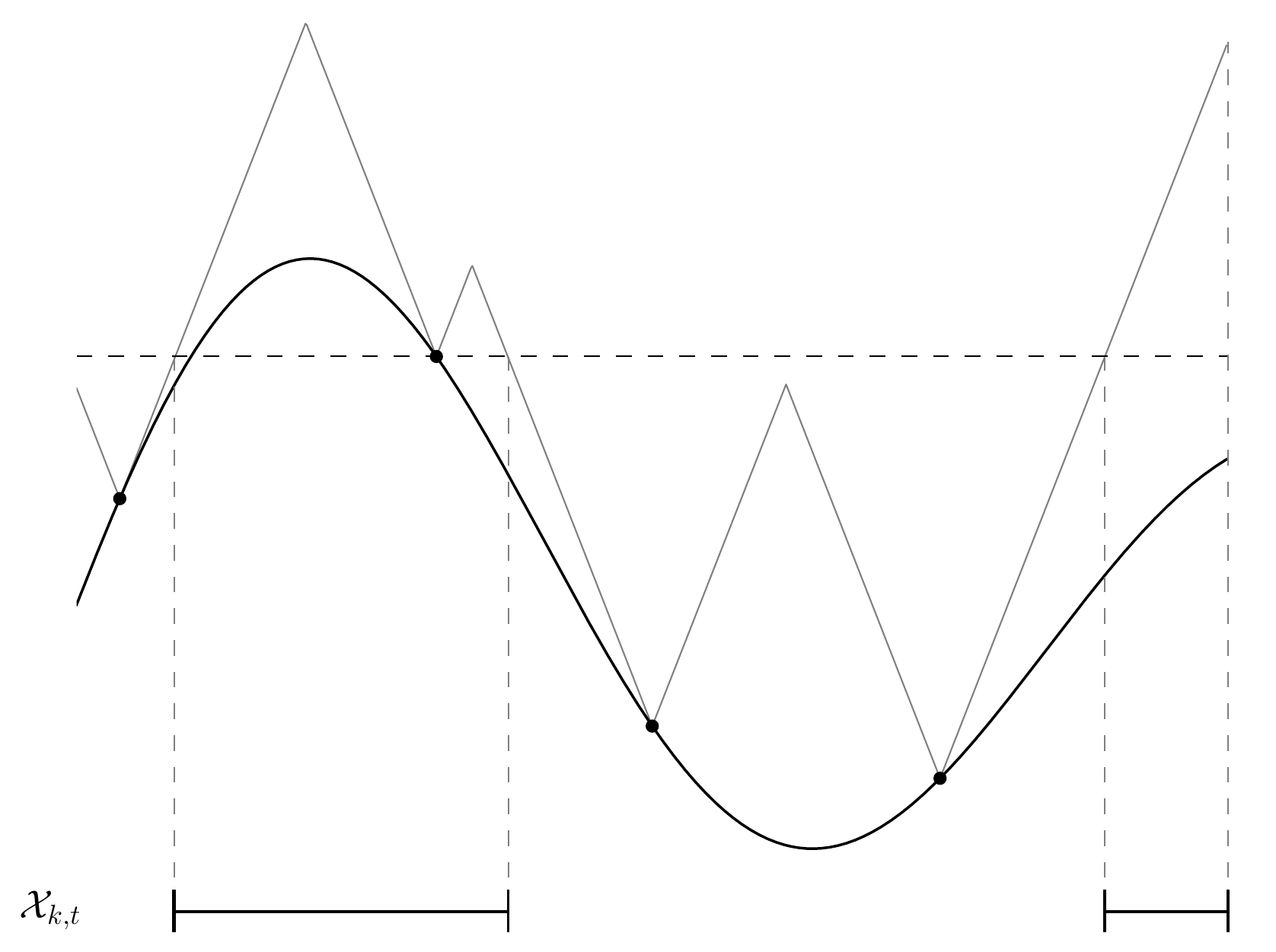}
\end{center}
\vspace{-1.5em}
\caption{From \cite{lipo}. (a) A simple Lipschitz function $f(x)$ evaluated over a sample of 4 points (dots). (b) The upper limit function, $U(x)$ is sketch in grey. (c) Here the domain of $f(x)$ is reduced to regions containing the maximum.}
\label{fig:lipoub}
\end{figure}

A piecewise linear upper bound, $U(x)$, of $f(x)$ is used to decide which $x$ to evaluate at each of the optimisation steps. Given already evaluated points $x_1, x_2, \ldots, x_j$,  $U(x)$ can be represented by:
\be
U(x) = \max_{i=1\dots j} \, \left( f(x_i) + k\,|x - x_i|\right)\,,
\ee
where $k$ is the Lipschitz constant for $f(x)$. By the definition of the Lipschitz constant, this will give $U(x) \geq f(x)$ for all $x$. The algorithm selects a test point, $x_t$ randomly, and then check if $U(x_t)$ is better than the best of the points so far chosen. If true, then $x_t$ is selected as the next point at which $f(x)$ should be evaluated. For illustration, Fig.~\ref{fig:lipoub} (a) shows a 4-point sample for a simple function $f(x)$. The grey lines in plot (b) show the upper bound function $U(x)$ constructed from the 4 sample points in (a). Plot (c) shows the region of $f(x)$ which satisfies the decision rule for selecting $x_t$ at which to evaluate $f(x)$ next. It can be seen that the procedure evolves such that over subsequent steps the selected points will eventually reach the global maximum of the $f(x)$. 

In practice the Lipschitz constant is not known, the functions to be minimised (known or unknown) can be very noisy, discontinuous or stochastic. LIPO could also be slow in converging to the maximum though it finds the local region near the maximum quickly. The implementation in DLIB is modified to tackle these problems. 

\section{A hybrid GA/Clustering/Amoeba algorithm to find all minima in a potential}
\label{AppD}

Finding all the stable or metastable minima in a system is a common problem in physics, but  heuristic search techniques tend to focus on the more iconic problem of correctly finding the global minimum. While individual algorithms may be well suited to this one task, collecting the locations of all the local minima as well as the global one in a given search space may be more efficiently done by combining them, as indeed suggested by the analysis in the previous appendix. In this appendix we describe an algorithm that is a combination of a Genetic Algorithm (GA) \cite{JH75,David89,JH92,Reeves02}, followed by a Cluster Algorithm (CA), followed by a Nelder-Mead (or amoeba) algorithm (NM) \cite{Nelder,lag1,lag2}. 

GAs seek optimal solutions by evolving a population of models in the search-space which, by means of a suitable definition of `fitness', is transformed into a fitness landscape. Such algorithms are able to avoid stagnation and attempt to find the global minima in NP-hard problems, and there has been some interest in their use in various contexts in particle physics, for example in \cite{Metcalfe:2000xn,Allanach:2004my,Mokiem:2005qf,Akrami:2009hp,Nesseris:2012tt,Blaback:2013ht,Damian:2013dq,Damian:2013dwa,Blaback:2013fca,Blaback:2013qza,Abel:2014xta,Hogan:2014qsa,Ruehle:2017mzq,Abel:2018ekz,Bull:2019cij}.  
However it is known that once a GA begins to select the favoured minimum the final stage of convergence is relatively slow. On the other hand  optimisation methods such as the Nelder-Mead algorithm flow to local minima and cannot address NP-hard problems, but in the basin of attraction of a minimum the NM method can converge much more quickly than a GA, if the function does not have many discontinuities and the dimensionality of the search-space is not too large. This suggests that a combination of these techniques may be beneficial. As we will see, for the problem at hand there are other benefits. 

\begin{figure}
\centering{}\includegraphics[scale=0.5]{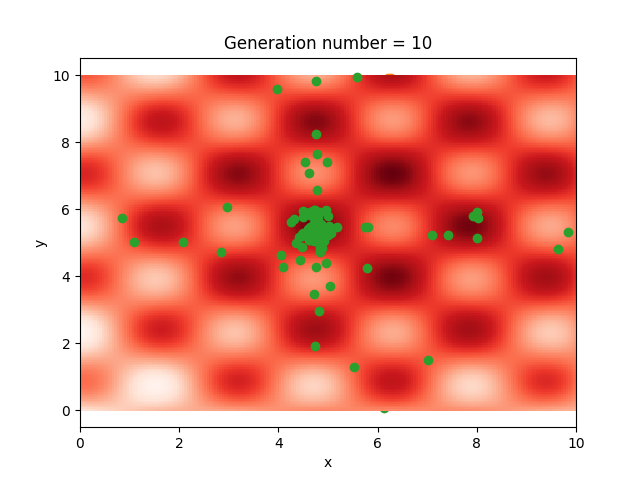}\includegraphics[scale=0.5]{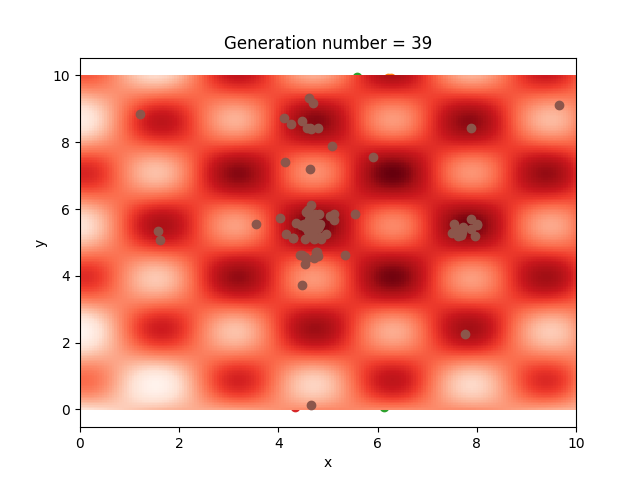}
\caption{\emph{Initial evolution of a genetic algorithm over a test function. There is rapid convergence to local clusters but further convergence takes place slowly. The population in this case is 100 individuals. \label{fig:evolution}}}
\end{figure}

In order to illustrate our procedure, we will consider finding the local minima for the test function:
\be
V (x,y) = - ( 36 \sin(2y) \cos(2x)+12(x+y) - x^2  - y^2 )\,.
\ee
This function can be seen plotted in the background of Fig. \ref{fig:evolution}. The evolution of a standard GA with 100 individuals is also plotted in the figure. Although many popular GAs are available, for our purposes it is more convenient to use our own tailored code. In particular in order to treat very flat potentials we use a simple roulette wheel selection genetic algorithm with fitness based on ranking (so the GA works for potentials with directions that are arbitrarily flat). As can be seen, within only a few generations the genetic algorithm clusters around the local minima. Continued convergence to the global minimum then proceeds slowly. It can be enhanced by dialing down the mutation rate (due to the relevant theorem of Holland) but then one risks losing all the information about the local minima. One approach to the problem of identifying all the local minima is to dial down the mutation `too early', so that sub-populations gets trapped in them. However it is much faster to pass to a NM algorithm as soon as clusters have formed. 
    
\begin{figure}
\centering{}\includegraphics[scale=0.5]{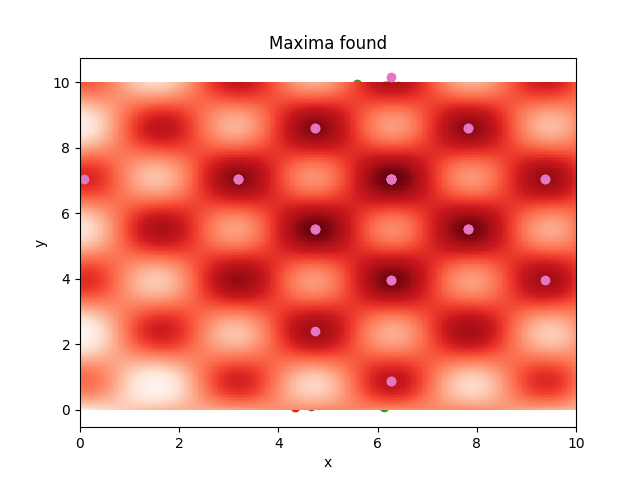}
\caption{\emph{Instead we run the GA several times for a few generations. Then a kmeans algorithm determines a set of clusters. A representative  $d+1 = 3$ dimensional simplex is chosen from each cluster, and a Nelder-Mead optimisation determines the local minimum (corresponding to the maximum fitness in the figure) for each simplex. \label{fig:evolution2}}}
\end{figure}

We therefore treat the problem in 3 stages. First we perform the genetic algorithm itself but with a relatively large (optimal) mutation rate. For the function we show here, a population of around 100 was found to be optimal. The process is terminated at an early stage after only a few generations, resulting in clusters located around the local minima as we observed in Fig. \ref{fig:evolution}. Note that if we wait too long, some of these clusters will begin to disappear unless we also impose some kind of crowding penalty. An additional interesting point is that (in contrast with a straight genetic algorithm) the overall process is better with larger populations as the clusters that are initially established are better defined. We then formally identify the clusters using a kmeans clustering algorithm. From each cluster we then select the best $d+1$ points (for a $d$ dimensional function) in order to form a representative $d+1$ dimensional simplex. Finally we determine the local minimum for each simplex using an NM (or amoeba) algorithm.  Overall the process is very fast for our example, and yields the outcome shown in Fig. \ref{fig:evolution2}. We include pseudo-code for the procedure in Algorithm \ref{alg}, and actual code is available at \url{www.maths.dur.ac.uk/~dma0saa/GANM/}. 
  
\begin{algorithm}[H]
 \KwData{Potential to be minimised over specified domain in $d$-dimensional space}
 \KwResult{Array of location and depth of local minima within domain }
 GA\;
 \For{population}{Initialise genotype}
 \While{generations $<$ gen-number}{\For{population}{ Find phenotype (potential) \\ Assign fitness by ranking}
 {\For{new-population} {Roulette wheel select breeding pair from {\it population}\\ Two-point crossover to create new individual \\ Elitist Mutation of new individual}
 }{{\it population} $=$ {\it new-population} }}
 Clustering: Run K-means clustering algorithm to produce  {\it clusters}\\
 {\it simplices = $ []$}\\
 \For{clusters}{Find phenotype\\ Assign fitness by ranking \\ Select first $d+1$ elements to form {\it simplex} \\ {\it simplices} += {\it simplex}  }
 
Nelder-Mead\;
{\it local-minima = $ []$}\\
 \For{simplices}{Find several {\it trial-minima} from {\it simplex} and perturbations\\ 
                       \For{trial-minima}{ test for unbounded below \\ test for flat directions and form basis\\
                                                     \If{trial-minimum passes {\bf \&\&} minimum {\bf $\notin$} local-minima}{{\it local-minima} += {\it trial-minimum}  }}}
  \caption{\label{alg} \it Pseudo-code for combined GA+Cluster+Amoeba algorithm}
\end{algorithm}
\mbox{ } \\
  \noindent   There are several other advantages since for cases of relatively low dimension and without discontinuities, the NM algorithm is known to converge rapidly and (like the GA itself) does not need a differentiable function \cite{lag1,lag2}. However the problem in this paper becomes 6D and some local minima appear close to the K\"ahler cone where there is a discontinuity. Thus the NM algorithm is somewhat less efficient even if the method still functions well.

\section{List of CY threefolds with $h^{1,1} = 2$}
\label{AppE}

\begin{table}[h!]
\centering
\begin{tabular}{|c||c|c|c|c|c|c|c|c|c|}
\hline
Model & $-\chi$ & $k_{111}$ & $k_{112}$ & $k_{122}$ & $k_{222}$ & K\"ahler cone & LVS & K3-fibred & Hard \\
\hline
$M_{2,1}$  & 54  & 0    & 1    & 1 & 0    & $t_1 > 0, \, \, t_2 > 0$       &  &  & \checkmark \\
$M_{2,2}$  & 72  & 9    & 0    & 0 & 9    & $t_1 < 0, \, \, t_1 + t_2 > 0$ & \checkmark &  & \\
$M_{2,3}$  & 144 & 1    & 0    & 0 & 3    & $t_1 < 0, \, \, t_1 + t_2 > 0$ & \checkmark &  & \\
$M_{2,4}$  & 144 & 1    & 0    & 0 & 3    & $t_1 < 0, \, \, t_1 + t_2 > 0$ & \checkmark &  & \\
$M_{2,5}$  & 144 & 1    & 0    & 0 & 3    & $t_1 < 0, \, \, t_1 + t_2 > 0$ & \checkmark &  & \\
$M_{2,6}$  & 162 & 0    & 3    & 3 & 0    & $t_1 > 0, \, \, t_2 > 0$       & & & \checkmark \\
$M_{2,7}$  & 164 & 2    & 0    & 0 & 5    & $t_1 < 0, \, \, t_1 + t_2 > 0$ & \checkmark &  & \\
$M_{2,8}$  & 168 & 0    & 0    & 4 & 8    & $t_1 > 0, \, \, t_2 > 0$       & & \checkmark & \\
$M_{2,9}$  & 168 & $-$4 & 4    & 0 & 0    & $t_1 > 0, \, \, t_2-t_1 > 0$   & & \checkmark & \\
$M_{2,10}$ & 168 & 0    & 0    & 4 & $-$4 & $t_1 - t_2 > 0, \, \, t_2 > 0$ & &  \checkmark & \\
$M_{2,11}$ & 168 & 2    & 4    & 0 & 0    & $t_1 > 0, \, \, t_2 > 0$       & & \checkmark & \\
$M_{2,12}$ & 168 & 0    & 0    & 4 & 5    & $t_1 > 0, \, \, t_2 > 0$       & &  \checkmark & \\
$M_{2,13}$ & 168 & 2    & 0    & 0 & 3    & $t_1 < 0, \, \, t_1 + t_2 > 0$ & \checkmark & & \\
$M_{2,14}$ & 168 & $-$1 & 0    & 4 & 11   & $t_1 < 0, \, \, t_1 + t_2 > 0$ & &  & \checkmark \\
$M_{2,15}$ & 168 & $-$1 & $-$2 & 0 & 5    & $t_1 < 0, \, \, t_1 + t_2 > 0$ & &  & \checkmark \\
$M_{2,16}$ & 168 & $-$1 & 0    & 4 & 14   & $t_1 < 0, \, \, t_1 + t_2 > 0$ & &  & \checkmark \\
$M_{2,17}$ & 168 & 8    & $-$2 & 0 & 6    & $t_1 < 0, \, \, 3t_1 +t_2 > 0$ & &  & \checkmark \\
$M_{2,18}$ & 176 & 3    & 0    & 0 & 5    & $t_1 < 0, \, \, t_1 + t_2 > 0$ & \checkmark &  & \\
$M_{2,19}$ & 180 & 3    & 0    & 0 & 3    & $t_1 < 0, \, \, t_1 + t_2 > 0$ & \checkmark &  & \\
$M_{2,20}$ & 186 & 8    & $-$2 & 0 & 14   & $t_1 < 0, \, t_1 + t_2 > 0$    &  &  & \checkmark \\
$M_{2,21}$ & 200 & 8    & 0    & 0 & 24   & $t_1 < 0, \, \, t_1 + t_2 > 0$ & \checkmark &  & \\
$M_{2,22}$ & 208 & 9    & 0    & 0 & 36   & $t_1 < 0, \, \, t_1 + t_2 > 0$ & \checkmark &  & \\
$M_{2,23}$ & 208 & 9    & 0    & 0 & 36   & $t_1 < 0, \, \, t_1 + t_2 > 0$ & \checkmark &  & \\
$M_{2,24}$ & 208 & 9    & 0    & 0 & 36   & $t_1 < 0, \, \, t_1 + t_2 > 0$ & \checkmark &  & \\
$M_{2,25}$ & 228 & 1    & 0    & 0 & 1    & $t_1 < 0, \, \, t_1 + t_2 > 0$ & \checkmark &  & \\
$M_{2,26}$ & 236 & 1    & 0    & 0 & 2    & $t_1 < 0, \, \, t_1 + t_2 > 0$ & \checkmark &  & \\
$M_{2,27}$ & 240 & 9    & 0    & 0 & 63   & $t_1 < 0, \, \, t_1 + t_2 > 0$ & \checkmark &  & \\
$M_{2,28}$ & 240 & 9    & 0    & 0 & 63   & $t_1 < 0, \, \, t_1 + t_2 > 0$ & \checkmark &  & \\
$M_{2,29}$ & 240 & 9    & 0    & 0 & 63   & $t_1 < 0, \, \, t_1 + t_2 > 0$ & \checkmark &  & \\
$M_{2,30}$ & 252 & 0    & 0    & 2 & 4    & $t_1 > 0, \, \, t_2 > 0$       & & \checkmark & \\
$M_{2,31}$ & 252 & $-$2 & 2    & 0 & 0    & $t_1 > 0, \, \, t_2-t_1 > 0$   &  & \checkmark & \\
$M_{2,32}$ & 252 & 0    & 0    & 2 & $-$2 & $t_1 - t_2 > 0, \, \, t_2 > 0$ &  & \checkmark & \\
$M_{2,33}$ & 252 & 0    & 0    & 2 & 0    & $t_1 > 0, \, \, t_2 > 0$       &  & \checkmark & \\
$M_{2,34}$ & 252 & $-$4 & 2    & 0 & 0    & $t_1 > 0, \, \, t_2-t_1 > 0$   &  & \checkmark & \\
$M_{2,35}$ & 252 & 2    & 0    & 0 & 1    & $t_1 < 0, \, \, 2t_1 + t_2 >0$ & \checkmark &  & \\
$M_{2,36}$ & 260 & 2    & 0    & 0 & 2    & $t_1 < 0, \, \, t_1 + t_2 > 0$ & \checkmark &  & \\
$M_{2,37}$ & 260 & 9    & 0    & 0 & 9    & $t_1 < 0, \, \, t_1 + t_2 > 0$ & \checkmark &  & \\
$M_{2,38}$ & 284 & 8    & 0    & 0 & 8    & $t_1 < 0, \, \, t_1 + t_2 > 0$ & \checkmark &  & \\
$M_{2,39}$ & 540 & 9    & 0    & 0 & 72   & $t_1 < 0, \, \, t_1 + 2 t_2>0$ & \checkmark &  & \\
\hline
\end{tabular}
\caption{Topological data for CYs with $h^{1,1}=2$ which are relevant to check the consistency of the extrema. Notice that $M_{2,17}$ and $M_{2,20}$ are `hard' but admit a non-diagonal dP divisor.}
\label{tab_cydata-h11eq2}
\end{table}

\section{List of structureless LVS with $h^{1,1} = 3$}
\label{AppF}

\begin{center}
\begin{tabular}{|c||c|c|c|c|c|c|c|} 
\hline
Model & $-\chi$ & $k_{111}$ & $k_{222}$ & $k_{223}$ & $k_{233}$ & $k_{333}$ & K\"ahler cone  \\
\hline
$M_{3,1}$  & 126 & 1 & 8  & $-$5 & 3 & 0 & $t_1 < 0, \,\, t_3 -2t_2 > 0, \,\, t_1 + t_3 > 0, \,\, t_1 + 3 t_2 >0$ \\
$M_{3,2}$  & 132 & 2 & $-$1 & 3 & $-$5 & 8 & $t_1 < 0, \, \, 3 t_3 -t_2 > 0, \,\, t_1 + t_3 > 0, \, \, t_2 - 2 t_3 >0$  \\
$M_{3,3}$  & 132 & 2 & $-$1 & 2 & 0 & $-$2 & $t_1 < 0, \, \, 2 t_3 -t_2 > 0, \,\, t_1 + t_3 > 0, \, \, t_2 - t_3 >0$ \\
$M_{3,4}$  & 138 & 3 & 0 & 3 & 3 & 0 & $t_1 < 0, \quad t_1 + t_2 > 0, \quad t_1 + t_3 > 0$ \\
$M_{3,5}$  & 144 & 3 & $-$1 & 1 & 3 & 0 &  $t_1 < 0, \quad t_1 + t_2 > 0, \quad t_3 -t_2> 0$ \\
$M_{3,6}$  & 144 & 3 & $-$1 & $-$1 & 3 & 0 & $t_1 < 0, \quad t_1 + t_2 > 0, \quad t_3 - t_2 > 0$ \\
$M_{3,7}$  & 144 & 3 & $-$1 & 2  & 0 & $-$2 & $t_1 < 0, \, \, 2 t_3-t_2 > 0, \,\, t_1 + t_3 > 0, \, \, t_2 - t_3 >0$ \\
$M_{3,8}$  & 144 & 3 & $-$1 & 3 & $-$5 & 8 &  $t_1 < 0, \, \, 3 t_3 -t_2 > 0,\,\, t_1 + t_3 > 0, \, \, t_2 - 2 t_3 >0$ \\
$M_{3,9}$  & 162 & 3 & 8 & $-$5 & 3 & 0 & $t_1 < 0, \quad t_1 + t_2 > 0, \quad t_3-2 t_2 > 0$ \\
$M_{3,10}$ & 164 & 8 & $-$2 & $-$2 & 6 & 14 & $t_1 < 0, \quad t_1 + t_2 > 0, \quad t_3- t_2 > 0$ \\
$M_{3,11}$ & 164 & 9 & $-$4 &  $-$2 & 8 & 22 & $t_1 < 0, \quad t_1 + t_2 > 0, \quad t_3- t_2 > 0$ \\
$M_{3,12}$ & 200 & 9 & $-$1 & $-$3 & 9 & 48 & $t_1 < 0, \quad t_1 + t_2 > 0, \quad t_3- t_2 > 0$ \\
$M_{3,13}$ & 216 & 9 & $-$1 & 1 & 17 & 64 & $t_1 < 0, \quad t_1 + t_2 > 0, \quad t_3- t_2 > 0$ \\
$M_{3,14}$ & 216 & 9 & $-$1 & 1 & 17 & 64 & $t_1 < 0, \quad t_1 + t_2 > 0, \quad t_3- t_2 > 0$ \\
\hline
\end{tabular}
\captionof{table}{List of structureless LVS models with $h^{1,1} = 3$.}
\label{tab_cydata-h11eq3-14-hard-LVS}
\end{center}

\newpage
\bibliography{reference}

\end{document}